\documentclass[%
longbibliography,
 nofootinbib,
 preprint,
superscriptaddress,
showpacs,preprintnumbers,
 amsmath,amssymb,
 aps,
pra,
]{revtex4-1}

\usepackage{graphicx, graphics}
\usepackage{dcolumn}
\usepackage{bm}
\usepackage{rotating}
\usepackage{ulem}		
\usepackage[percent]{overpic}
\usepackage{color}
\newcommand{\red}[1]{\textcolor{black}{#1}}
\usepackage[autostyle=true]{csquotes}

\begin{document}

\preprint{APS/123-QED}

\title{Angular momentum transport and flow organisation in Taylor-Couette flow at radius ratio of $\bm{\eta=0.357}$}

\author{Andreas Froitzheim}%

\author{Sebastian Merbold}%

\affiliation{Brandenburg University of Technology Cottbus-Senftenberg, Department of Aerodynamics and Fluid Mechanics, Siemens-Halske-Ring 14, D-03046, Cottbus, Germany }
\author{Rodolfo Ostilla-M\'{o}nico}%
\affiliation{Cullen College of Engineering, University of Houston, Houston, TX 77204, USA}

\author{Christoph Egbers}%
\email{Christoph.Egbers@b-tu.de}
\affiliation{Brandenburg University of Technology Cottbus-Senftenberg, Department of Aerodynamics and Fluid Mechanics, Siemens-Halske-Ring 14, D-03046, Cottbus, Germany }

\date{\today}

\begin{abstract}
We experimentally and numerically investigate the angular momentum transport in turbulent Taylor-Couette flow for independently rotating cylinders at a small radius ratio of $\eta=0.357$ for various shear Reynolds numbers ($4.5\times 10^3 \leq Re_S \leq 1.2 \times 10^5$) and ratios of angular velocities ($-0.5 \leq \mu \leq 0.2$). \red{The momentum transport in terms of the pseudo-Nusselt number ${Nu}_\omega$ does not show a pure power law scaling with the forcing $Re_S$ and features non-constant effective scaling between $1.3\times 10^4 \leq Re_S \leq 4 \times 10^4$. This transition lies in the classical turbulent regime and is caused by the curvature-dependent limited capacity of the outer cylinder to emit small-scale plumes at a sufficient rate to equalize the angular momentum in the bulk.} For counter-rotating cylinders, a maximum in the torque occurs at $\mu_{\max}=-0.123 \pm 0.030$. The origin of this maximum can be attributed to a strengthening of turbulent Taylor vortices, which is revealed by the flow visualization technique. In addition, different flow states at $\mu_{\max}$ concerning the wavelength of the large-scale vortices have been detected. The experimental and numerical results for the Nusselt number show a very good agreement. 
\end{abstract}

\maketitle


\section{\label{sec:Introduction}Introduction}

Complex turbulent flows are often investigated in simple geometries to reveal fundamental mechanisms of transport and structure formation. For rotating shear flows, the Taylor-Couette (TC) geometry, which consists of coaxial and independently rotating cylinders enclosing the working fluid in between, is a widely used configuration to study hydrodynamic instabilities \citep{DiPrima1985} as well as turbulence \citep{Grossmann2016}. Next to the shear forces and rotation forces, curvature effects also play an important role in Taylor-Couette flow. The influence of curvature can be measured in terms of the curvature number $R_C=d/\sqrt{r_1r_2}$ (cf. \cite{Dub2005}), which is the reciprocal of the geometrical mean of both cylinder radii ($r_1,r_2$) normalized by the width of the gap $d$ and measures the importance of the additional coupling between wall-normal (radial) and streamwise (azimuthal) velocities induced by the curvilinear coordinates. This coupling causes strong modifications in shear flows. Indeed, \citet{Bradshaw1973} noted \enquote*{\textit{the surprisingly large effect exerted on shear-flow turbulence by curvature of the streamlines in the plane of the mean shear}}, when studying boundary layers over curved surfaces. The curvature number can also be expressed in terms of the radius ratio $\eta=r_1/r_2$. As $\eta$ approaches to 1, $R_C$ becomes zero and \red{curvature} effects vanish. In this situation, Taylor-Couette flow becomes rotating plane Couette flow \cite{Brauckmann2016}. The TC flow is also in close analogy to the Rayleigh-B\'enard \red{(RB)} \red{convection} \cite{Eckhardt07a,Eckhardt07b}, the fluid flow in a layer heated from below and cooled from above, where curvature is absent. As shown by \citet{Brauckmann2016}, when $\eta <0.9$ curvature effects become important and result in differences between the inner and outer gap region concerning their linear stability, intermittency and fluctuations. This asymmetry inside the gap increases, when $\eta$ decreases. \red{It is worth to mention, that such kind of asymmetry can be established also in a RB flow between the top and bottom plate, if the adjusted temperature difference is sufficiently high. In that case the temperature-dependent fluid properties are no longer neglectable and so-called Non-Oberbeck-Boussinesq effects come into play \citep{Ahlers2006}.} Thus, the investigation of wide-gap TC flows, where curvature effects gain more importance are of special interest.

The geometry of a Taylor-Couette setup is defined by the radius ratio $\eta=r_1/r_2$, the gap width $d=r_2-r_1$ and the aspect ratio $\Gamma=\ell /d$ with the radii of the inner $r_1$ and outer cylinder $r_2$ and their length $\ell $ (cf. Fig. \ref{fig:sketch}(a)). When the angular velocities $\omega_{1,2}$ of both cylinders are normalized using the geometrical parameters and the kinematic viscosity of the fluid $\nu$, two Reynolds numbers $Re_{1,2}=(r_{1,2}\omega_{1,2}d)/\nu$ can be defined, also known as the inner and outer cylinder Reynolds number. The shear driving of TC flow is given by the differential rotation of the cylinders, i.e. the difference in velocities between inner and outer cylinder. This can be  non-dimensionally represented as a shear Reynolds number $Re_S$ \cite{Dub2005}:

\begin{equation}
Re_S=\dfrac{2}{1+\eta} \vert \eta Re_2-Re_1 \vert=\dfrac{u_S d}{\nu}.
\end{equation}

\noindent The shear Reynolds number results from non-dimensionalizing a characteristic shear velocity $u_s$ obtained from examining TC flow in a rotating frame of reference, where the amount of rotation $\Omega_{rf}$ is chosen such that the speed of cylinders is equal but with opposite sign. A further control parameter which controls the magnitude of the Coriolis forces in the rotating reference frame is required to completely characterize the system. This can be either denoted as a rotation number $R_\Omega=2\Omega_{rf} d/U_S $, or, more intuitively, using the ratio of angular velocities $\mu=\omega_2/\omega_1$, where a positive sign denotes co-rotation and a negative sign counter-rotation. When a flow state is defined by prescribing the values of $Re_S$ and $\mu$, the global system response can be quantified by the transport of angular momentum $J_\omega$ \citep{Eckhardt07a,Eckhardt07b}, which is preserved over the radial coordinate $r$:

\begin{equation}
J_\omega=r^3 \left( \left\langle u_r \omega \right\rangle_{\varphi,z,t} -\nu \partial_r \left\langle \omega \right\rangle_{\varphi,z,t} \right).
\end{equation}

\noindent The brackets $\left\langle \cdot \right\rangle_{\varphi,z,t}$ denote an average over a cylindrical surface and time and $u_r$ and $\omega$ denote the radial and angular velocity, respectively. \red{$J_\omega$ is directly proportional to the torque $\mathcal{T}$ that is necessary to apply at the cylinders to keep them at a constant speed}: {$J_\omega=\mathcal{T}/(2\pi \ell \rho)$}. When the momentum transport is normalized by its corresponding laminar value ${J_{\omega}^{lam}=2\nu r_1^2r_2^2(\omega_1-\omega_2)/(r_2^2-r_1^2)}$, we obtain a pseudo-Nusselt number ${Nu}_\omega$, analogous to the Nusselt number which is used to quantify the heat flux in e.g. the Rayleigh-B\'{e}nard flow:

\begin{equation}
{Nu}_\omega=\dfrac{J_\omega}{J_\omega^{lam}}.
\end{equation}

There are three main questions related to the momentum transport in TC flows which have been typically addressed: (\emph{i}) what are the \red{effective} scaling laws relating momentum transport and driving, expressed non-dimensionally as a Reynolds number ($Re_1$ or $Re_S$), (\emph{ii}) how does the momentum transport depend on the rotation ratio $\mu$ and (\emph{iii}) how do the large-scale structures known as \enquote*{\textit{Taylor vortices}} influence this transport. \red{Typically, the \red{effective} scaling of the momentum transport is analyzed expressed in terms of the dimensionless torque $G=\mathcal{T}/(2\pi \ell \rho \nu^2)$ for pure inner cylinder rotation ($\mu=0$) by assuming a power law ansatz of the form $G\sim Re_{1,S}^\alpha$, which translates for the pseudo-Nusselt number to ${Nu}_\omega \sim Re_{1,S}^{\alpha-1}$. Besides, to emphasize the similarities between the TC flow and RB convection, also the effective scaling of ${Nu}_\omega$ with the so-called Taylor number $Ta=\sigma^2 Re_S^2$ is used, where $\sigma=2^{-4}(1+\eta)^4\eta^{-2}$ denotes the pseudo-Prandtl number, leading to ${Nu}_\omega \sim Ta^{\frac{\alpha}{2}-\frac{1}{2}}$}. By using marginal stability analysis, \citet{King1984} and \citet{Marcus1984} predicted \red{effective} scalings for the torque in the spirit of what had been done for Rayleigh-B\'{e}nard flow by \citet{Malkus1958}. Marginal stability theory is based on a separation of the flow into an inner boundary layer (BL), a bulk flow and an outer boundary layer. Both BLs are assumed to be laminar and the bulk flow as well as the BLs are marginally stable according to Rayleigh's stability criterion. As consequence, the radial profile of the angular momentum \red{$L=\omega r^2$} has to be flat in the bulk flow, which is in good agreement with the findings of previous studies \citep{Smith1982, Dong07, Brauckmann2016}. Matching of the $L$-profiles at the separation borders and independence of the momentum transport on the radial coordinate leads to a predicted scaling exponent of $\alpha=5/3$ identical to the one for Rayleigh-B\'{e}nard flow and the calculations of \citet{Barcilon1984}. Another prediction for the \red{effective} torque scaling was derived by \citet{Lathrop92} using a Kolmogorov type argument assuming that the energy dissipation rate $\epsilon$ is constant in the inertial range and has no length scale dependence. This condition yields a sort of upper bound for the scaling exponent and is equivalent to the assumption of an infinite Reynolds number, where viscous effects can be neglected. In contrast to Lathrop, we use the shear velocity $u_S$ as the characteristic velocity difference and the gap width $d$ as the characteristic length, leading to the relation

\begin{align}  \label{equ:predic_lath}
\epsilon &=\dfrac{2}{r_2^2-r_1^2}\left( \omega_1 - \omega_2 \right) \nu^2 G \sim \dfrac{u_S^3}{d^3},\\
G & \sim \dfrac{\eta}{(1-\eta)^2} Re_S^2,
\end{align}

\noindent with $\alpha=2$. The same exponent was also found by \citet{Doering1992} directly derived from the Navier-Stokes equation as an upper bound. Further, \citet{Eckhardt07b} deduced a prediction for the \red{effective} torque scaling in analogy to the RB flow by distinguishing the contributions of the bulk and the BL to the overall transport. For not-too-wide gaps ($\sigma \approx 1$) and laminar boundary layers, they found a scaling exponent of $\alpha=5/3$, when the boundary layers are dominant and $\alpha=2$, when the bulk flow is dominant. Contrary to these predictions, no pure power law scaling could be detected within numerous numerical and experimental studies. \citet{Wendt1933} performed direct torque measurements for three radius ratios ($\eta=0.68, \ 0.85$, and $0.935$), finding $\alpha=1.5$ for $4\times 10^2 \leq Re_1 \leq 10^4$ and $\alpha=1.7$ for $Re_1>10^4$. Instead of a partially constant exponent, \citet{Lathrop92} identified a monotonically increasing exponent from $\alpha=1.23-1.87$ in the range of $8\times 10^2 \leq Re_1 \leq 1.2 \times 10^6$ and a sharp discontinuity in the $\alpha$-$Re_1$-curve around $Re_1=1.3\times 10^4$ for $\eta=0.724$. In addition, \citet{Merbold13} measured the torque for $\eta=0.5$, where $\alpha$ was increasing up to $Re_S=6\times 10^5$, before it settled to a nearly constant value around $\alpha=1.65 \pm 0.03$. Again, a transitional behavior of the exponent was found around $Re_S\approx 8\times 10^4$-$10^5$. Similar results for a wide range of radius ratios of $\eta= 0.714, \ 0.769, \ 0.833$ and $0.909$ with a universal scaling exponent of $\alpha=1.78$ for Taylor numbers larger than $Ta=\sigma^2 Re_S^2>10^{10}$ were found by \citep{Ostilla14a}. The only investigation of the momentum transport at very wide-gap TC flows for a radius ratio of $\eta=0.35$ was performed by \citet{Burin2010} finding $\alpha=1.6 \pm 0.1$ for $2\times 10^3 \leq Re_1 \leq 10^4$ and $\alpha=1.77 \pm 0.07$ for $2\times 10^4 \leq Re_1 \leq 2\times 10^5$. There, $J_\omega$ was measured only locally and indirect by Laser Doppler Velocimetry (LDV). \red{We want to stress that the last mentioned result contradicts the previous investigations, where a monotonic increase of the transitional Reynolds number with decreasing $\eta$ were reported. In the study of \citet{Burin2010} however, the bulk dominated regime ($\alpha > 5/3$) is reached at $Re_1\approx 2\times 10^4$ much earlier than in the case of $\eta=0.5$}. Besides, in all presented investigations a transitional behavior was found in the scaling exponent, whose origin was linked by \citet{Ostilla14b} to a shear instability in the BLs, where laminar BLs would transition to turbulent ones. This picture unifies the aforementioned scaling predictions. At low Reynolds numbers, the momentum transport is limited by the laminar boundary layers ($\alpha=5/3$), while at large Reynolds numbers, the featureless turbulent bulk is the limiting factor ($\alpha=2$). 

When independently rotating cylinders come into play, the torque exhibits an additional dependency on the rotation ratio $\mu$ and it can be shown, that a separation ansatz of the form $G=f_1(\mu)\cdot f_2(Re)$ is valid \citep{Dub2005,Paoletti2012}. Furthermore, the torque depicts a maximum in the slight counter-rotating regime, when the shear Reynolds number is kept constant and only the rotation rate $\mu$ is changed. This maximum location strongly depends on the radius ratio and was found in various numerical and experimental studies ($\mu_{\max}(\eta=0.5)=-0.195$ and $\mu_{\max}(\eta=0.71)=-0.361$ by \citet{Brauckmann13b}, $\mu_{\max}(\eta=0.5)=-0.198$ by \citet{Merbold13}, $\mu_{\max}(\eta=0.714,0.716)=-0.33$ by \citet{VanGils2012} and $\mu_{\max}(\eta=0.724)=-0.33$ by \citet{Paoletti11}). Note, that we excluded studies for radius ratios $\eta>0.8$ due to the fact, that the flow behavior changes as described by \citet{Brauckmann2016}. A first attempt to explain the physical mechanism behind this maximum was made by \citet{VanGils2012} with the so-called angle bisector hypothesis. \citet{VanGils2012} argued that the most unstable point in the parameter space is the location equally distant from both Rayleigh stability lines $\mu=\eta^2$ and $\mu=\infty$, which defines the rotation ratio of the torque maximum $\mu_b$ by:

\begin{equation}
\mu_b=\dfrac{-\eta}{\tan\left[ \dfrac{\pi}{2}-\dfrac{1}{2}\arctan\left( \eta^{-1} \right) \right]}.
\end{equation} 

\noindent While the predicted rotation ratio of the torque maximum for $\eta=0.716$ becomes $\mu_b=0.368$ in good agreement with the measurements of \citet{VanGils2012} of $\mu_{\max}=-0.33$, it deviates noticeable for $\eta=0.5$, where it becomes $\mu_b=-0.309$ in contrast to the measurements of \citet{Merbold13} with $\mu_{\max}=-0.198$.

Another prediction, linking the location of the torque maximum to the onset of intermittency in the gap and a strengthening of large-scale Taylor vortices, was developed by \citet{Brauckmann13b}. Their theory states that turbulent Taylor vortices in the slight counter-rotating regime can extend the theoretical neutral line by \red{a factor $a(\eta)$:}

\begin{equation}
\red{a(\eta)=(1-\eta)\left[\sqrt{\dfrac{(1+\eta)^3}{2(1+3\eta)}}-\eta \right]^{-1}}.
\end{equation}

\noindent \red{$a(\eta)$ is called parameter of vortex extension and depends only slightly on $\eta$, reaching values in the range of $1.4-1.6$ \citep{Esser1996}.} If this extension exactly ends at the outer cylinder wall, the Taylor vortices are most pronounced leading to a maximum in transport. For even higher counter-rotating rates, the vortices are restricted to an inner gap region with a laminarized outer region. As the momentum transport has to be constant over the whole gap, intermittent bursts flushing from the unstable inner gap region into the stable outer gap region appear. Based on the stability calculations of \citet{Esser1996}, the prediction of \citet{Brauckmann13b} can be written as:

\begin{equation} \label{equ:pred_brauck}
\mu_p(\eta)=-\eta^2\dfrac{(a^2(\eta)-2a(\eta)+1)\eta+a^2(\eta)-1}{(2a(\eta)-1)\eta+1},
\end{equation}

\noindent with $a(\eta=0.357)=1.532$. A connection between the torque maximum and the onset of intermittency was also identified via LDV-measurements by \citet{VanGils2012} and the strengthening of TV has been shown by \citet{Ostilla14b} and \citet{Froitzheim2017}. In comparison, the prediction yields $\mu_p(\eta=0.5)=-0.191$ and $\mu_p(\eta=0.71)=-0.344$ in very good agreement with the before mentioned findings. This is a clear indication that Taylor vortices play a prominent role in the momentum transport in the fully turbulent regime for the \enquote*{\textit{broad peak}} identified by \citet{Brauckmann2016}. Furthermore, the wavelength or the total number of turbulent Taylor vortices inside the gap influences the amount of momentum transport \citep{Brauckmann13a,Martinez2014}. In line with these findings, \citet{Huisman2014} and \citet{vanderVeen2016} showed, that multiple vortex states can exist in the gap for counter rotation at very high Taylor numbers of $Ta=10^{13}$ and that these states are stable leading to different amounts of momentum transport depending on the actual flow state.
Based on this state of the art, we analyze the scaling of the Nusselt number with the shear Reynolds number and the location of the torque maximum by means of direct torque measurements and numerical simulations for the radius ratio of $\eta=0.357$. Furthermore, we perform flow visualisations to determine the corresponding flow state itself.

\section{Experimental setup and measurement technique} \label{sec:exp}

\begin{figure}[htb]
\centering
\includegraphics[trim=0 0 170 0, clip]{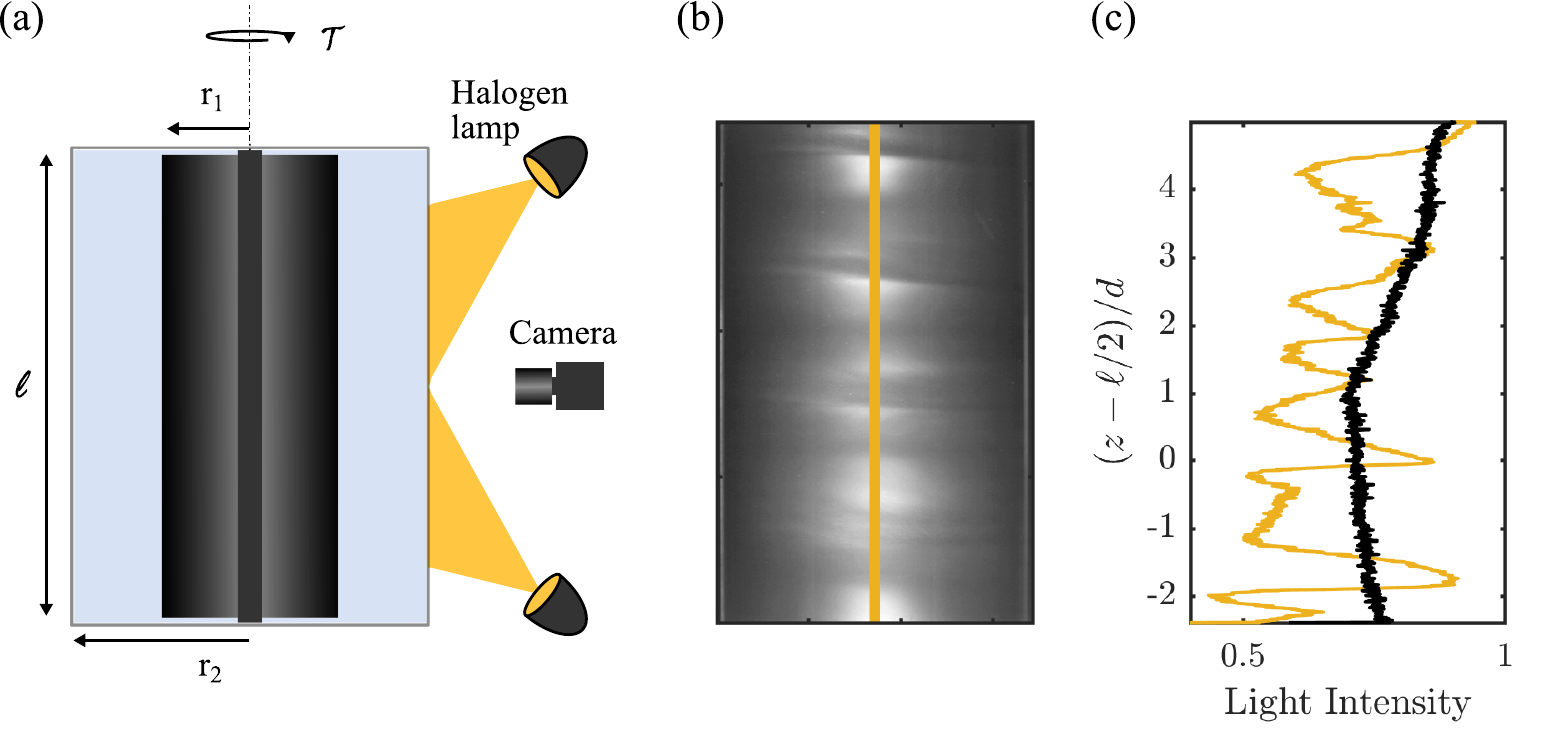}
\caption{(a) Sketch of TC apparatus with visualisation setup. (b) Time-averaged image over 2900 images of the flow for $Re_S=3\times 10^3$ and $\mu=0$ with an acquisition frequency of 60\,Hz. Yellow line indicates position of evaluation of axial light intensity profiles. (c) Yellow line represents light intensity profile along the line of (b). Black line indicates light intensity profile for an laminar flow, taken to correct for inhomogeneous illumination.}
\label{fig:sketch}
\end{figure}

The experiments have been carried out in the Top view Taylor-Couette Cottbus facility (TvTCC), which has been already used in \cite{Froitzheim2017}. The apparatus consists of an aluminum anodized inner cylinder (IC) and a transparent, acrylic glass outer cylinder (OC) with radii $r_1=25\,\textrm{mm}$ and $r_2=70\,\textrm{mm}$, respectively. Their length is $\ell=700\,\textrm{mm}$, which results in a gap width of $d=45\,\textrm{mm}$, a radius ratio of $\eta=0.357$ and an aspect ratio of $\Gamma=15.6$. Also the top plate is made of transparent acrylic glass and both end plates rotate together with the OC. The inner cylinder is mounted on an inner, stainless steel shaft, that is connected to the driving of the IC. In the IC driving train, a contactless shaft to shaft rotary torque sensor (DR2500/M220-G21) of the company Lorentz Messtechnik GmbH with a nominal torque of $\pm 2\,\textrm{Nm}$ and an accuracy of $\pm 0.1\%$ is integrated via torsionally stiff couplings with a transmissible torque of $6.17\,\textrm{Nm}$. Both cylinders can rotate independently up to $n_1=\pm 2000\, \textrm{rpm}$ and $n_2=\pm 500\, \textrm{rpm}$ with a standard deviation smaller than 1\%. As working fluids, the silicone oils $\text{M20}$, $\text{M10}$, $\text{M5}$ and $\text{M3}$ are used and the temperature of the fluid is monitored with a thermocouple type K sensor inserted into the bottom end of the gap, whose measurement accuracy is 0.8\%. This temperature is used to calculate the actual fluid viscosity and density. The temperature dependent fluid data, measured with a Stabinger viscosimeter with integrated density measurement cell (Anton Paar, SVM 3000/G2, $\Delta \nu= \pm 0.35\, \% $, $\Delta \rho = \pm 0.0005\,\text{g/cm}^3$), are shown in Table\,\ref{tab:fluid}. Note, that the temperature can only be measured for an outer cylinder at rest. 

\begin{table}
\caption{\label{tab:fluid} Kinematic viscosity $\nu$ and density $\rho$ of the working fluids as function of the temperature $\text{T}$. Functions are calculated based on a linear regression of the viscosimetry data. Correlation coefficients of regression are larger than 0.99.}
\begin{ruledtabular}
\begin{tabular}{lcccccc}
 & \multicolumn{3}{c}{kinematic viscosity $\nu\,[\text{mm}^2\text{/s}]$} \footnote{$\nu(T\,[^\circ C])=aT+b$} & \multicolumn{3}{c}{density $\rho\,[\text{g/cm}^3]$}\footnote{$\rho(T\,[^\circ C])=cT+d$} \\ \hline
  & a & b& @ 25$^\circ \text{C}$ & c & d& @ 25$^\circ \text{C}$ \\ \hline
  M3 & $-0.046$ & $4.175$ &3.023& $-9.356\times 10^{-4}$ & $0.913$&0.890\\
  M5 & $-0.088$ & $7.612$ &5.402& $-9.247\times 10^{-4}$ & $0.939$&0.916\\ 
  M10 & $-0.189$ & $15.05$ &10.34& $-8.952\times 10^{-4}$ & $0.958$ & 0.935\\
  M20 & $-0.385$ & $29.57$ &19.95& $-8.963\times 10^{-4}$ & $0.972$&0.950 
\end{tabular}
\end{ruledtabular}
\end{table}

According to the presented setup, the torque is measured over the whole length of the inner cylinder including end wall effects and the friction of three stainless steel single-row groove ball bearings. Therefore, the torque is measured twice, once with a fluid filled gap and once with air. Afterwards, both torque signals are subtracted to get the torque induced by pure friction force of the fluid. It is worth to mention, that measurement points have been rejected, where the torque signal using air exceeds 20\% of the value with liquid. The torque sensor setup is validated using a larger inner cylinder with $r_{1,val}=35\,mm$, leading to a radius ratio of $\eta_{val}=0.5$. In this configuration, the torque has been measured for $Re_S=5\times 10^3$ and $Re_S=10^4$ in the range of $\mu \in [-0.6;0]$ and compared with direct torque measurements and direct numerical simulations of \cite{Merbold13}. We find a very good agreement with a relative deviation of smaller than 5\% \cite{Merb2018}. For the torque measurements of the current study at $\eta=0.357$, a fixed protocol was used. We can distinguish between measurements for $\mu=0$, where the IC speed is increased sequentially, and measurements at a fixed shear Reynolds number $Re_S$, where the ratio of angular velocities is changed sequentially. In both cases, the cylinder speeds of the first measurement point are set in with an subsequent waiting time of approximately 10 minutes to let the flow evolve. Afterwards, the torque is measured for 90 seconds at 10\,Hz before the cylinder speeds are adapted for the next measurement point. Here, the required change in cylinder speed is small and the waiting time can be reduced to 120 seconds. Due to this waiting time, we can assume the measurements to be quasi-stationary. In case of pure inner cylinder rotation, the fluid temperature is measured for each data point directly before the measurement phase. In case of differential rotating flow states, the inner cylinder speed is adapted always before the outer one and for every fourth data point, the outer cylinder speed is set to zero and the fluid temperature is measured.

For the purpose of flow visualisation, two halogen lamps are mounted on the experiment stand at the height of about the top and bottom plate, respectively and centered to the rotation axis. Both lamps are tilted into the direction of mid height of the experiment to illuminate an axially middle segment of the flow. Further, an Optronis CR 3000x2 high speed camera with $1690\times 1710\, \textrm{px}$ sitting on a tripod was positioned in front of the experiment capturing approximately 6 gap widths in axial direction and the whole outer cylinder diameter in radial direction centered slightly above mid height. To make the fluid motion visible the silicone oils are suspended with \red{anisotropic} aluminum flake particles, which have an elliptic shape and an upper diameter of approximately $5\, \mu \textrm{m}$. \red{As shown experimentally by \citet{Abcha2008} using Kalliroscope flakes in a TC system, \enquote*{\textit{small anisotropic particles align with the flow streamlines by giving the precision on the velocity component which bears these alignment.}} In their setup, they used a light sheet visualisation in the radial-axial plane, which led to an intensity image of the reflected light that corresponds to the magnitude of the radial velocity component in good agreement with numerical simulations of \citet{Gauthier1998} For our setup, where the flow is visualised in the azimuthal-axial plane, the intensity image should provide information on the azimuthal $u_\varphi$ or axial velocity component $u_z$, meaning that bright areas are related to large absolute values of $u_\varphi$ or $u_z$.} Moreover, blue pigments (BASF, Heliogen blue, K6850) have been added into the fluid to reduce its transparency and focus on the flow in the outer gap region. This is necessary, as we will visualize fully turbulent flows in a wide-gap configuration. When the experiment is not running, the aluminum and pigment particles settle down slowly. After approximately 3 days, all particles have been mixed out and gather at the bottom plate. However, a uniformity of suspension can be restored within a few minutes by setting up a fully turbulent flow state. For each visualised flow state, a waiting time of 10 minutes was used before images of the flow were taken. Each visualisation consists of 2900 images recorded at a frequency of 60\,Hz. From the acquired images, only one central line (yellow line in Fig.\,\ref{fig:sketch}b) is analyzed over time to reveal the flow organisation. Regarding inhomgeneous illumination, the intensity distribution $\mathcal{I}_0$ for a laminar flow (black line in Fig.\,\ref{fig:sketch}c) is subtracted from the original image intensity $\mathcal{I}$ (yellow line in Fig.\,\ref{fig:sketch}c).

\section{Numerical Simulations}

To complement the experiments, direct numerical simulations of Taylor-Couette flow are performed by solving the incompressible Navier-Stokes equations in cylindrical coordinates in a rotating reference frame:

\begin{equation}
 \displaystyle\frac{\partial \textbf{u}}{\partial t} + \textbf{u}\cdot \nabla \textbf{u} + 2\Omega_{rf}\times\textbf{u} = -\nabla p + \nu \nabla^2 \textbf{u},
 \end{equation}
 
 \begin{equation}
  \nabla \cdot \textbf{u} = 0,
 \end{equation}

\noindent where $\textbf{u}$ is the velocity, $p$ the pressure, $\Omega_{rf}$ the angular velocity of the rotating frame and $t$ is time. Spatial discretization is achieved by the use of a second-order energy-conserving centered finite difference scheme, and time-marching is performed with a low-storage third-order Runge-Kutta for the explicit terms and a second-order Adams-Bashworth scheme for the implicit treatment of the wall-normal viscous terms. Further details of the algorithm can be found in \citep{ver96,poe15}. This code has been previously used and validated extensively for Taylor-Couette flow \citep{Ostilla14b}. 

Unlike the experiments, the numerical simulations use axially periodic boundary conditions with a periodicity length $L_z$. This is expressed non-dimensionally as an aspect ratio $\Gamma=L_z/d$, and is fixed to $\Gamma=2$. \red{For this aspect ratio, exactly one vortex pair with wavelength $\lambda_z/d=2$ fits into the simulated volume. Due to the periodic boundary conditions, this corresponds to the solution for infinitely long cylinders as long as periodicity effects are negligible. Indeed, it is known that one single roll is sufficient to accurately capture most statistics, including torque \citep{Brauckmann13a,Ostilla2016b}.  The non-dimensional radius ratio $\eta$ is also matched to that of the experiment $\eta=0.357$. Unlike previous studies, no additional rotational symmetry is imposed, and the full $2\pi$ azimuthal extent of the domain is simulated. Due to the wide-gap, the streamwise extent $L_x$ is already quite small at the inner cylinder as compared to the gap size: $L_x(r=r_1)/d \approx 3.5$. From our experience, a minimum of $L_x/d=\pi$ is needed to produce accurate decorrelations \cite{Ostilla2015}, so we suspect that introducing a degree of rotational symmetry which further reduces this extent will introduce artifacts.} The simulations are performed in the reference frame of \citet{Dub2005} such that both cylinders rotate with opposite velocities $\pm U/2$. In this frame the two control parameters naturally become the shear Reynolds number and the rotation parameter defined in the introduction. 

The shear Reynolds number is varied between $Re_s=5\times 10^3$ and $Re_s=4\times 10^4$, with resolutions ranging from $N_\theta \times N_r \times N_z=256\times256\times512$ to $768\times384\times768$, in the azimuthal, radial and axial directions respectively. The axial and azimuthal grid distributions are homogeneous, but points are clustered in the radial direction. Due to the asymmetry of the cylinders, higher resolutions are needed at the inner cylinder to properly resolve the structures. For the radial direction, this is solved by clustering more points. The azimuthal discretization becomes larger with increasing radius, which naturally provides for higher resolution in the inner cylinder. The axial direction can become problematic, as it is discretized in a homogeneous manner. In this direction, the resolution requirements are set by the structures at the inner cylinder, and thus the outer cylinder wall is overresolved in this direction. At low Reynolds numbers the resolution is chosen such that dispersive effects cannot be observed when looking at the flowfield. For higher Reynolds numbers, we measure the resolution in viscous units, i.e.~normalized by $\delta_\nu=\nu/u_\tau$, where $u_\tau$ the frictional velocity is $u_\tau=\sqrt{\tau_W/\rho}$, with $\tau_W=\mathcal{T}/(2 \pi r^2\ell)$ the shear at the wall. We choose the viscous unit at the inner wall as it is more restrictive due to the higher shear $\tau_W$. The spatial discretization is then approximately $r\Delta\theta^+ \approx 10$, $\Delta z^+\approx4$ and $\Delta r^+\in(0.5,5)$ in inner wall units, as in \cite{Ostilla2016b}.

The rotation parameter $R_\Omega$ is varied between $R_\Omega\in(0.08,0.84)$, corresponding to values of $\mu\in(-0.3,0.075)$. Table \ref{tab:dnsresults} in the appendix contains full details of the resolutions used. Temporal convergence is assessed by measuring the difference in torque between both cylinders, and ensuring that the time-average of both torques coincides within $3\%$. 

\section{Scaling of the Nusselt number with the shear Reynolds number for pure inner cylinder rotation}

\begin{figure}[htb]
\centering
\includegraphics{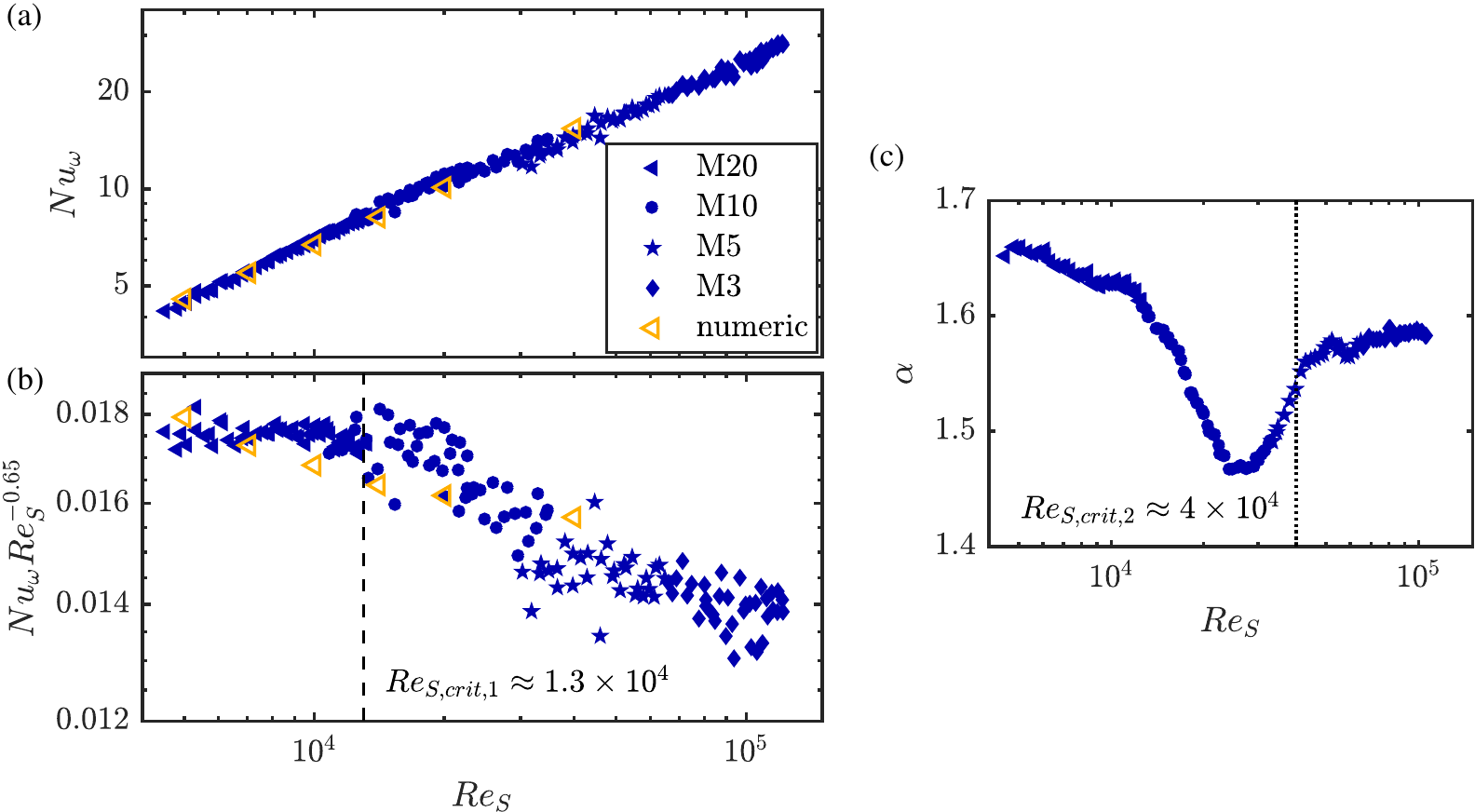}
\caption{(a) The dependence of the normalized torque in terms of the Nusselt number $Nu_\omega$ on the shear Reynolds number $Re_S$ for pure inner cylinder rotation $\mu=0$ in a range of $4.5\times 10^3 \leq Re_S \leq 1.2\times 10^5$. Different filled blue symbols represent experimental data for different working fluids. Yellow open symbols denote numerical results. (b) Dependence of the Nusselt number compensated by $Re_S^{-0.65}$ as function of the shear Reynolds number. A change in the dependence appears at approximately $Re_{S,crit,1} \approx 1.3 \times 10^4$, which is marked as dashed line. (c) Corresponding local scaling exponent $\alpha$ as function of the shear Reynolds number calculated for a bin size of $\Delta_{10}(Re_S)=0.5$. A transitional behavior is visible in the region $1.3\times 10^4 \leq Re_S \leq 4 \times 10^4$. The dotted line indicates the end of this transition at $Re_{S,crit,2}\approx 4 \times 10^4$, \red{where $\alpha$ starts to monotonically increase.}}
\label{fig:torque_scale}
\end{figure}

We first start by studying the \red{effective} scaling of the torque with Reynolds number. The torque for pure inner cylinder rotation is measured over a range of $4.5\times 10^3 \leq Re_S \leq 1.2\times 10^5$ using several working fluids. In Fig.\,\ref{fig:torque_scale}a we show the non-dimensionalized torque, i.e.~the Nusselt number as function of the shear Reynolds number for all used working fluids in logarithmic scale. The experimental data points across working fluids only scatter slightly in a small area without observable discontinuities at the crossovers. In addition, the numerical data points are in good agreement with the experimental results. As could be expected, the non-dimensional momentum transport increases with an increasing shear Reynolds number. We first compensate the Nusselt number by $Re_S^{-0.65}$. This scaling exponent was found for $\eta=0.5$ by \citet{Merbold13} in the same Reynolds number range as this study (see Fig. \ref{fig:torque_scale}b). The compensated Nusselt number is approximately constant up to $Re_S\leq 1.3 \times 10^4$ and decreases for higher shear Reynolds numbers. This behavior suggests that some kind of transition takes place\red{,} changing the overall momentum scaling.
\newpage

\noindent To precisely characterize the transition, we calculate the local exponent $\alpha$ of the power law ansatz \red{${Nu}_\omega \sim Re_{S}^{\alpha-1}$}. As described by \citet{Lathrop92}, $\alpha$ is defined as:

\begin{equation}
\alpha(Re_S)=\dfrac{\partial \left( \log_{10} {Nu}_\omega \right)}{\partial \left( \log_{10} Re_S \right)}+1.
\end{equation}

\noindent Instead of directly calculating the derivative of the ${Nu}_\omega$-$Re_S$-curve, which may lead to difficulties due to the mentioned slight scattering of the data points and the existence of more than one data point at a specific Reynolds number, we compute a linear least square fit across equidistant intervals $\Delta_{10}(Re_S)$ in logarithmic scale centered by the individual points. The scaling exponent obtained by taking an interval of $\Delta_{10}(Re_S)=0.5$ is shown in Fig.\,\ref{fig:torque_scale}c. The exponent \red{$\alpha$} has a slow downwards trend starting at \red{$\alpha \approx 1.65$} up to $Re_S \approx 1.3 \times 10^4$, then strongly decreases to values around \red{$\alpha \approx 1.47$}, before it subsequently increases again noticeably. For shear Reynolds numbers above $Re_S>4 \times 10^4$, the exponent is monotonically increasing depicting a slight slope. Apparently, the \red{effective} momentum scaling reveals a transition confined by two critical Reynolds numbers, which are $Re_{S,crit,1} \approx 1.3 \times 10^4$ and $Re_{S,crit,2} \approx 4 \times 10^4$. \red{As the value of $\alpha$ is always smaller than $5/3$, the here observed transition cannot be connected to the transition from the classical to the ultimate regime, which coincidentally happens at a very similar value for the shear Reynolds number in the case of $\eta\geq 0.714$ ($Re_{S,crit}=1.04\times 10^4$ or $Ta_{crit}=3\times 10^8$) \citep{Ostilla14b}. It is further worth to mention, that the values of $\alpha$ calculated within this study are much smaller than the one found by \citet{Burin2010} for nearly the same $\eta$. They reported scaling exponents of $\alpha>5/3$ for $Re_1 \geq 2\times 10^4$ on the basis of local LDV measurements. This discrepancy, and in particular the question of whether our measurements reach ultimate regime or not, will be discussed in the further course.}

\begin{figure}[htb]
\centering
\includegraphics[trim=0 10 0 250, clip]{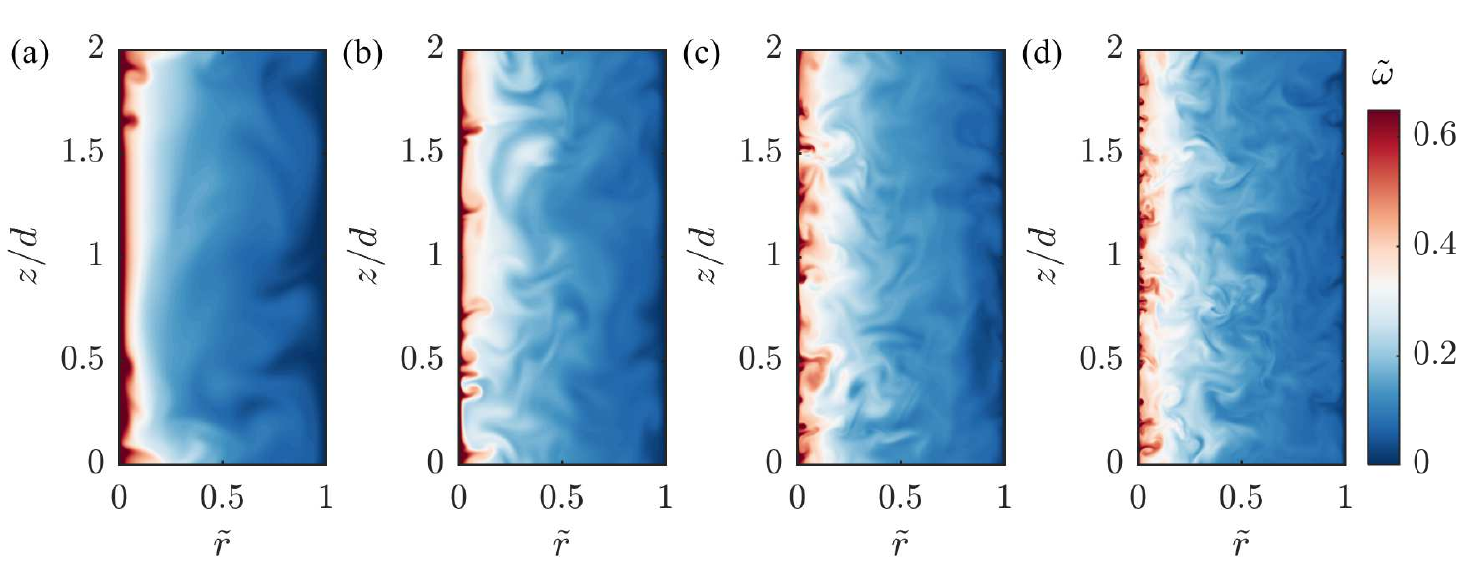}
\caption{Constant-azimuth cuts of normalized instantaneous angular velocity $\tilde{\omega}=(\omega-\omega_2)/(\omega_1-\omega_2)$ for four different $Re_S$ at $\mu=0$. From left to right: $Re_S=5\times 10^3$, $Re_S=1\times10^4$, $Re_S=2\times10^4$ and $Re_S=4\times10^4$. Data are based on numerical simulations.}
\label{fig:instL}
\end{figure}

Using the numerical simulations, we can explore the flow field to uncover other ways the flow changes during the transitions. We first show in Fig. \ref{fig:instL} constant-azimuth cuts of the instantaneous angular \red{velocity} for increasing shear Reynolds number. We can see that the structures at the inner cylinder appear to be smaller than the ones at the outer cylinder, as can be expected by the higher frictional Reynolds number ($Re_{\tau,o}=u_{\tau,o}d/(2\nu)=\eta Re_{\tau,i}$).  \red{Aside from the quiescent zone in the region of the inner cylinder for $Re_S=5\times 10^3$, which disappears at higher $Re_S$, no fundamental changes are occurring. The structures (plumes) of angular velocity become smaller and more numerous with increasing $Re_S$ as could be expected. We remind the reader that in a similar manner, no large changes in flow topology were observed for $\eta=0.5$ during the transitions by \cite{Ostilla14b}.}

\red{Focusing on the boundary layer, in Table \ref{tab:re_tau}, we summarize the frictional Reynolds numbers of our numerical simulations for $\mu=0$ at the inner and outer cylinder. For comparison purposes, we also include the values of $Re_\tau$ for the appearance of the ultimate regime happens for $\eta=0.5$, $\eta=0.714$ and $\eta=0.909$ (taken from Refs.~\citep{Ostilla14b,Ostilla14c}). This appearance is triggered by a boundary layer transition, and we could thus expect similar frictional Reynolds numbers for the transition.}

\begin{table}[htb]
\caption{\label{tab:re_tau} Overview of inner and outer cylinder frictional Reynolds number $Re_{\tau,i/o}$ for pure inner cylinder rotation in the transitional regime of this study and comparison with more narrow-gap investigations in the region of the classical-ultimate turbulent transition at $\eta=0.5$, $0.909$  from \citep{Ostilla14b} and $\eta=0.714$ from \citep{Ostilla14c}.}
\begin{tabular}{ccccc}
  \hline
  $\eta$ & $\quad \quad Re_S \quad \quad $ & $\quad \mu \quad$ & $\quad Re_{\tau,i} \quad$ & $Re_{\tau,o}$\\
  \hline
  0.357 & $5\times 10^3$ & 0 & 120 & 43\\
  0.357 & $1\times 10^4$ & 0 & 209 & 75\\
  0.357 & $2\times 10^4$ & 0 & 363 & 130\\
  0.357 & $4\times 10^4$ & 0 & 635 & 230\\
  0.5 & $7.9\times 10^4$ & 0 & 1112 & 556\\
  0.714 & $1.4\times10^4$ & 0 & 252 & 180\\
  0.909 & $1.4\times10^4$ & 0 & 240 & 220\\
    \hline

\end{tabular}
\end{table}

\red{From the table it can be seen that there is a markedly increased value of $Re_\tau$ for the transition at $\eta=0.5$. This was previously attributed to the effect of concavity in stabilizing the outer cylinder boundary layer \cite{Ostilla14c}. Accordingly, the here observed transition for $\eta=0.357$ in the range of $1.3\times 10^4\leq Re_S \leq 4\times 10^4$ is unlikely caused by the development of turbulent BLs, especially at the outer cylinder, due to the too low shear and curvature stabilization. Something to note, however, is that the inner frictional Reynolds number $Re_{\tau,i}$ at $Re_{S,crit,1}\approx 1.3\times 10^4$ has a similar value as $Re_{\tau,o}$ at $Re_{S,crit,2}\approx 4\times 10^4$, which can be interpreted as follows. The observed transition takes place at a frictional Reynolds number of about $Re_{\tau,crit}\approx 230$. As this value is reached earlier at the inner than at the outer cylinder, some transition happens in two steps. These results show the pronounced asymmetry inside such a wide-gap TC flow, which is probably the reason for the strong discrepancy between our globally measured effective scaling exponent and the locally measured one by \citet{Burin2010}.} 

\begin{figure}[htb]
\centering
\includegraphics{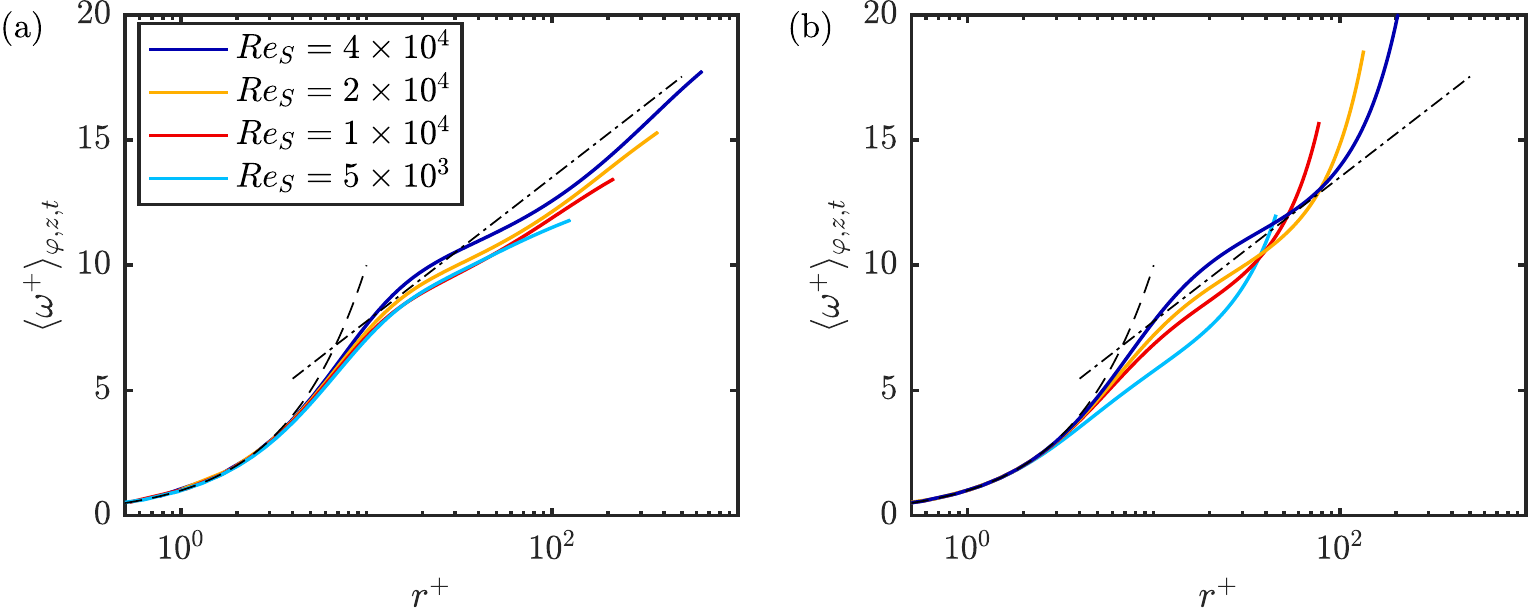}
\caption{Azimuthally, axially and temporally averaged normalized angular velocity profiles of (a) $\omega_i^+(r^+)$ near the inner ($\tilde{r} \in [0,0.5]$) and (b) $\omega_o^+(r^+)$ near the outer cylinder ($\tilde{r} \in [0.5,1]$). The angular velocities are defined as $\omega^+_i=(\omega(r_1)-\omega(r))/(u_{\tau,i}/r_1)$ and $\omega^+_o=(\omega(r)-\omega(r_2))/(u_{\tau,o}/r_2)$, while the radial coordinates are calculated by $r^+_i=(r-r_1)/\delta_{\nu,i}$ and $r^+_o=(r_2-r)/\delta_{\nu,o}$. The subscripts $i,o$ are omitted in the labels. The dashed line represents the viscous sublayer $\omega^+=r^+$ and the dash-dotted line a logarithmic law of the form $\omega^+=2.5 \ln(r^+)+2$ as guides for the eye.}
\label{fig:aveL_BL}
\end{figure}

If the transition we are seeing is not the transition to the ultimate regime, because it is inconsistent with the scaling law, then what is it? To further investigate the nature of the boundary layers that have been formed, the profiles of angular velocity are shown in Fig. \ref{fig:aveL_BL} for the inner and outer boundary layer in wall units. While the inner cylinder boundary layer could be showing some logarithmic behaviour with deviations due to curvature effects, the outer cylinder boundary is definitely of laminar type. 

\begin{figure}
\centering
\includegraphics{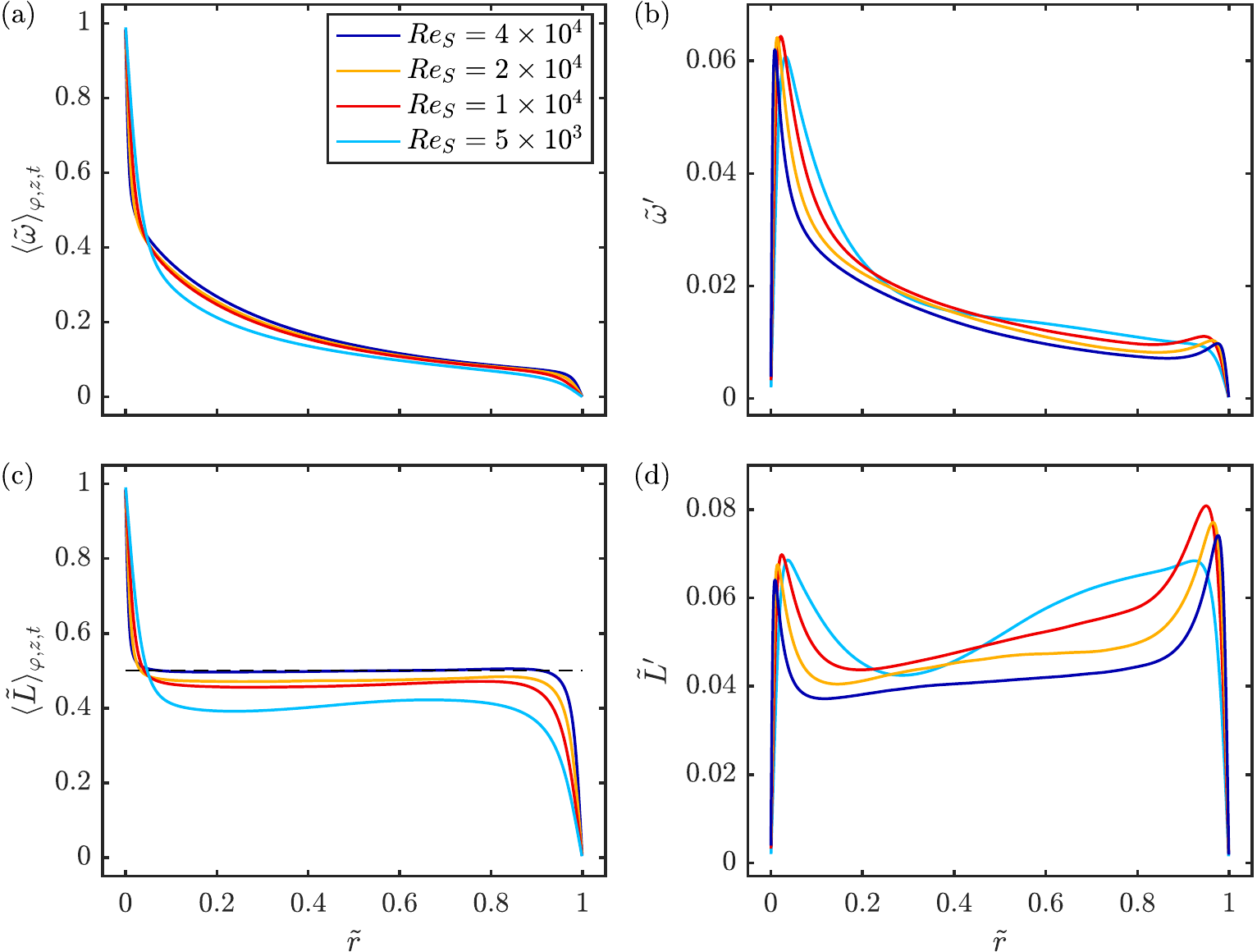}
\caption{Azimuthally, axially and temporally averaged normalized (a) angular velocity $\langle \tilde{\omega} \rangle_{\varphi,z,t}$ and (c) angular momentum $\langle \tilde{L} \rangle_{\varphi,z,t}$ for $\mu=0$ and varying $Re_S$. The corresponding root-mean-squared fluctuations of the normalized (b) angular velocity $\tilde{\omega}^\prime$ and (d) angular momentum $\tilde{L}^\prime$ are depicted on the right-hand side. Data are based on numerical simulations.}
\label{fig:aveL}
\end{figure}

The asymmetry between boundary layers is further explored in Fig. \ref{fig:aveL}, where we show the average and root-mean-squared profiles of angular velocity and angular momentum. The differences of the angular velocity profiles with increasing shear Reynolds number are not very visible, so we focus our analysis here on the angular momentum. There, it can be seen that only for the highest $Re_S$ achieved in the simulations the mean angular momentum does become equal to the arithmetic average $0.5$ (Fig. \ref{fig:aveL}c). This provides an indication that the transition seen at $Re_{S,crit,2}$ is related to the outer cylinder's capacity to emit plumes at a rate that achieves marginal stability and equalizes the angular momentum in the bulk. We note that not only is the shear smaller at the outer cylinder due to fact that the radius is larger (see Table \ref{tab:re_tau}), and thus the velocity gradient must be smaller to achieve the same torque, but that the concave curvature of the outer cylinder stabilizes the emission of turbulent plumes. Thus the shear required for plume emission becomes higher. Further evidence of this is seen in Fig. \ref{fig:aveL}d, where the root-mean-squared fluctuations of angular momentum are shown. For $Re_S=5\times10^3$, only a flat plateau at the outer cylinder is existing, indicating that there is no considerable production of fluctuations by the outer cylinder. Once $Re_S$ increases, a sharp peak of fluctuations at the outer cylinder boundary layer appears, probably coinciding with the transition seen at $Re_{S,crit,1}$, even if the number is slightly different for the simulations. Even once the sharp peak arises it appears that the turbulence level is not enough to equalize the angular momentum until $Re_{S,crit,2}$. Thus, we can suggest that the transitions in the scaling laws are associated to the capacity of the outer cylinder to emit angular momentum plumes to develop a marginally stable profile.

\section{Torque maximum}

\begin{figure}[htb]
\centering
\includegraphics{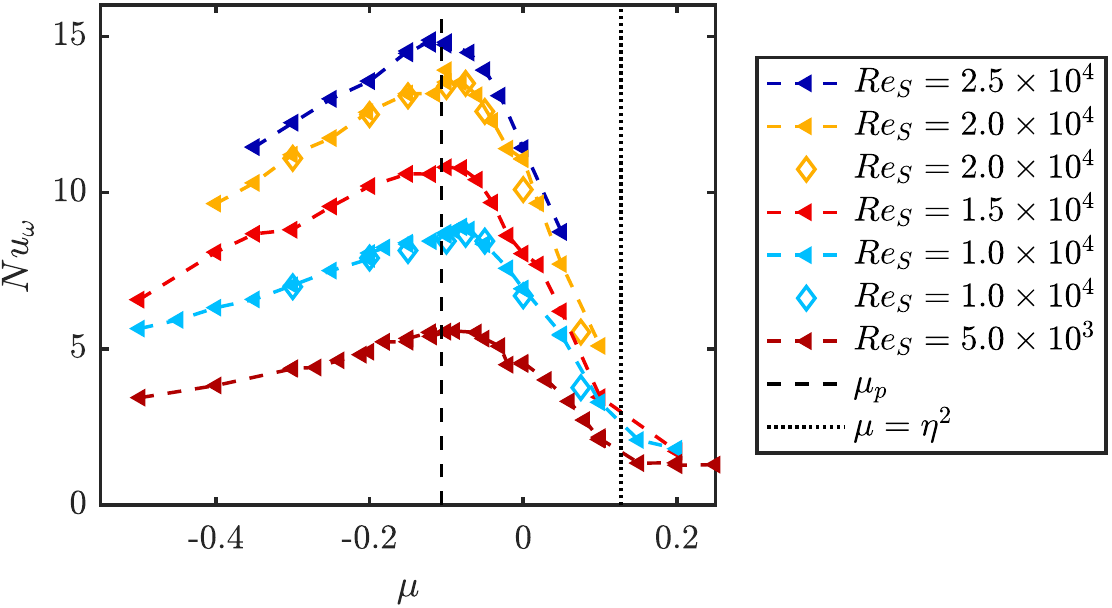}
\caption{Dependence of the Nusselt number $Nu_\omega$ on the rotation ratio $\mu$ for different shear Reynolds numbers. Filled symbols represent experimental data (already discussed in \cite{Merb2018}) and open symbols numerical data. The dashed line indicates the prediction of the torque maximum according to \citet{Brauckmann13b}, while the dotted line indicates the Rayleigh stability criterium.}
\label{fig:torque_mus1}
\end{figure}

We now turn to the dependence of the Nusselt number on the rotation ratio $\mu$, when the shear Reynolds number is kept constant. In Fig.\,\ref{fig:torque_mus1}, we show the combined numerical and experimental results in the range $5\times10^3<Re_s<2.5\times10^4$.

\noindent The Nusselt number exhibits a maximum in the low counter-rotating regime for all Reynolds numbers which asymptotically becomes closer to the prediction of \citet{Brauckmann13b}. The maximum location according to the angle bisector hypothesis ($\mu\approx -0.25$) lies noticeable far from the peak within the stronger counter-rotating regime. For $Re_S=10^4$ and $Re_S=2\times 10^4$, where also simulations have been performed, the experimental and numerical data agree very well. The increase of the Nusselt number coming from co-rotating cylinders to $\mu_{\max}$ is much steeper, than the decline for higher counter rotation. This behavior results probably from the small distance of $\mu_{\max}$ from the Rayleigh stability line at $\mu=\eta^2=0.128$ and will be revisited later. In the Rayleigh stability region, the flow should be laminar and the Nusselt number has to be 1. In our case, we find for the lowest investigated Reynolds number of $Re_S=5\times 10^3$ a nearly constant value of ${Nu}_\omega\approx 1.3$ in the Rayleigh stable regime. This value, which is slightly higher than expected, may be the result of end wall effects causing a large-scale circulation, the so-called Ekman vortices. Such a secondary flow can grow, especially in laminar flows, and fill the whole gap \citep{Avila2008}. 

\begin{figure}[htb]
\centering
\includegraphics{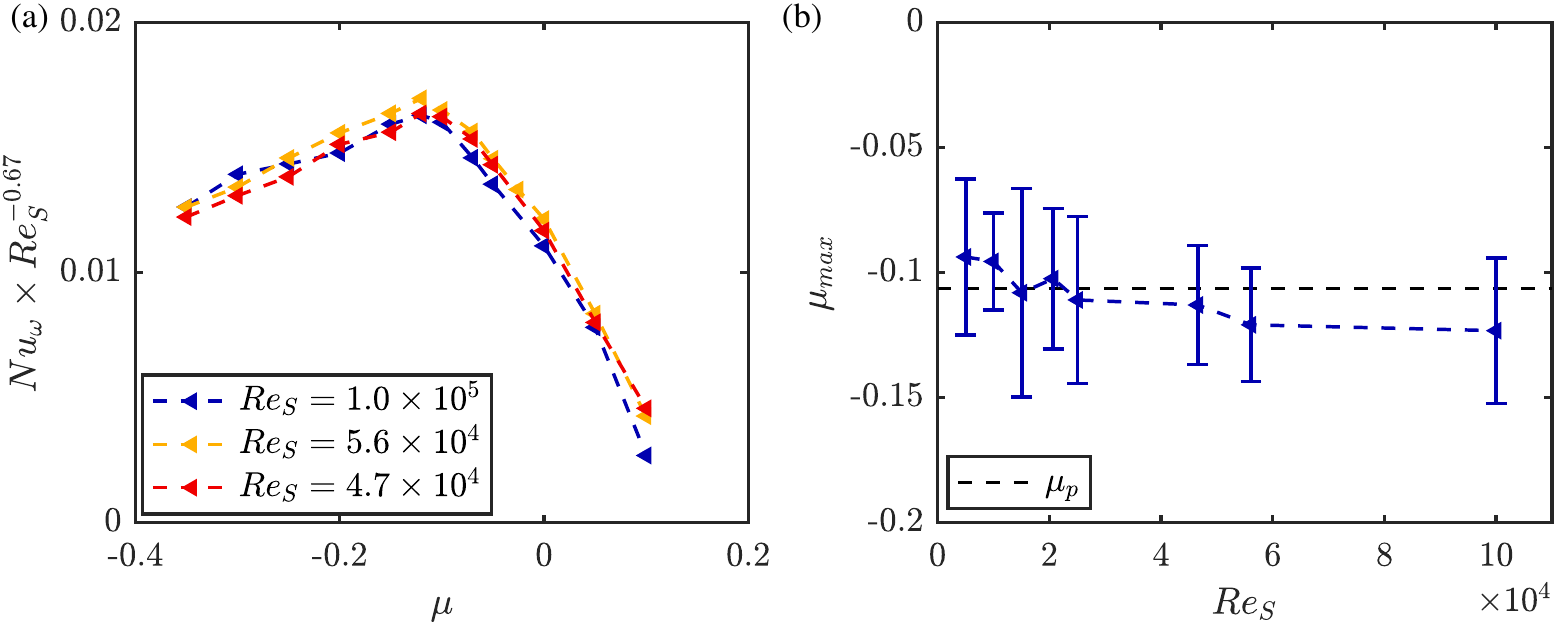}
\caption{(a) Dependence of the experimentally determined compensated Nusselt number ${Nu_\omega \times Re_S^{-0.67}}$ on the rotation ratio $\mu$ for three shear Reynolds numbers. (b) Evolution of the torque maximum location as function of the shear Reynolds number, calculated by a quadratic fit to the measured profiles in the range of $ -0.16 \leq \mu \leq -0.06$. The dashed line represents the predictions of the torque maximum according to \citet{Brauckmann13b}.}
\label{fig:torque_mus2}
\end{figure}

At higher shear Reynolds numbers, investigated only experimentally, the overall shape of the $Nu_\omega$ curve stays the same with a torque maximum close to $\mu_p$, see Fig.\,\ref{fig:torque_mus2}a. To determine the exact location of the torque maximum, we fit a quadratic curve of the form ${{Nu}_\omega=p_1 \mu^2+p_2 \mu +p_3}$ to the measured profiles in the range of ${-0.16 \leq \mu \leq -0.06}$. This way the location of the torque maximum is given by ${\mu_{\max}=-p_2/(2p_1)}$ and its uncertainty is defined according to \cite{Brauckmann13b} as


\begin{equation}
\Delta \mu_{\max}=\sqrt{\dfrac{\Delta \mathcal{T}}{p_1 \mathcal{T}} \left( Nu_{\omega}\right)_{\max}}.
\end{equation}

\noindent The location of $\mu_{\max}$ shifts with increasing Reynolds number to more negative values and appears to settle down to a nearly fixed value for $Re_S\geq 5.6\times 10^4$, reaching ${\mu_{\max}=-0.123 \pm 0.030}$ at $Re_S=10^5$ (see Fig.\,\ref{fig:torque_mus2}b). The prediction of \citet{Brauckmann13b} is within the uncertainty range of our maximum location.

\begin{figure}[htb]
\centering
\includegraphics{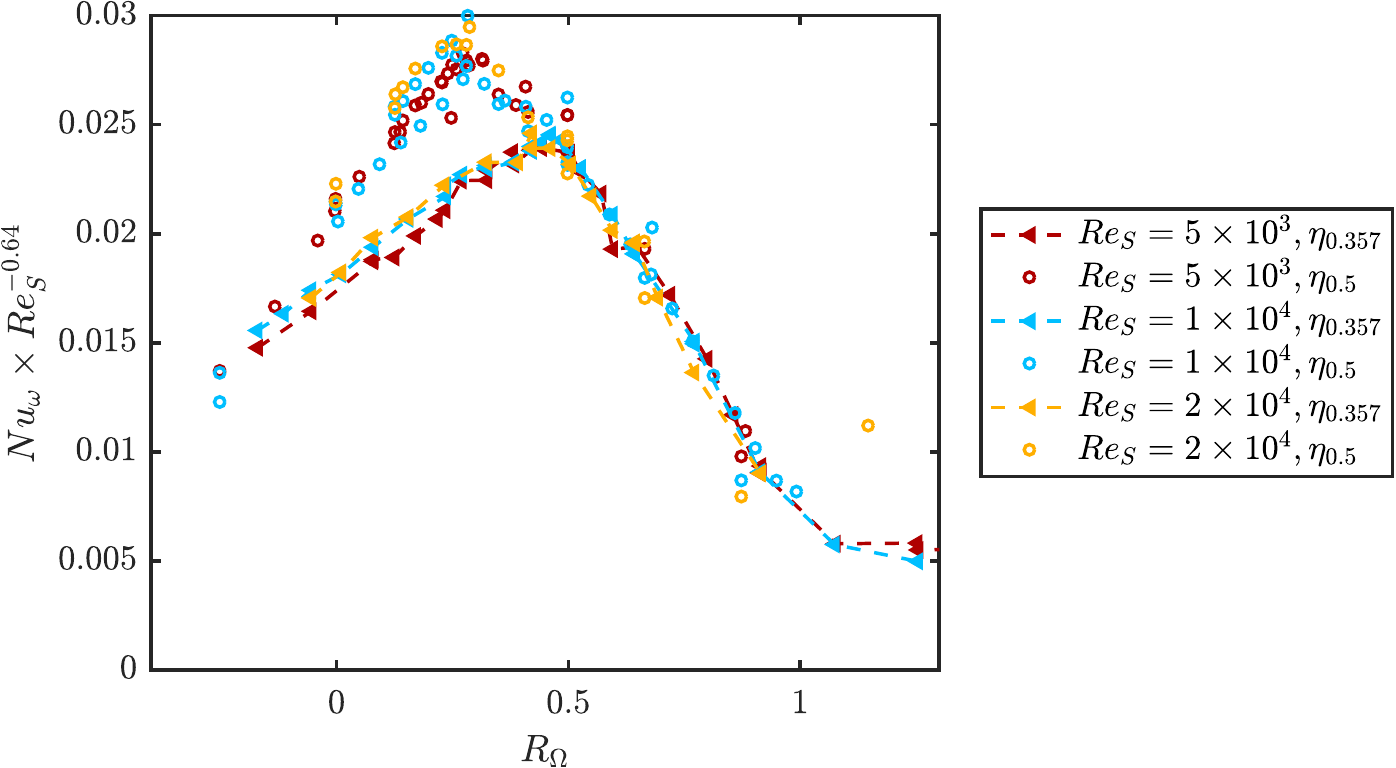}
\caption{Non-dimensional torque $Nu_\omega$ compensated by $Re_S^{-0.64}$ for three $Re_S$ and two wide-gap curvatures $\eta=0.357$ (experiments from this study) and $\eta=0.5$ (taken from \citep{Merbold13}) in terms of the rotation parameter $R_\Omega$. A collapse for the torques can be seen in the region between the Rayleigh stability line ($R_\Omega=1$) up to the appearance of radial partitioning, which happens at a lower $R_\Omega$ for $\eta=0.357$. }
\label{fig:vgl_maximum}
\end{figure}

To finalize this section, we show the compensated torque in terms of $R_\Omega$ vs $Nu_\omega \times Re_S^{-0.64}$. \citet{Dub2005} had already noted that the non-dimensional torque depends very little on the radius ratio $\eta$, and \citet{Brauckmann2016} further developed this by showing a wide collapse of $Nu_\omega(R_\Omega)$ for $R_\Omega>0.25$ for radius ratios higher than $\eta=0.5$, and for $R_\Omega>0.1$ for radius ratios higher than $0.8$. \citet{Brauckmann2016} noted that the region where the large-gap Nusselt number collapses is where it is not affected by 
the radial flow partitioning. The appearance of radial flow partitioning introduces a strong dependence on the radius ratio of the Nusselt numbers. Further proof of this for the wide-gap considered here can be seen in Fig. \ref{fig:vgl_maximum}. The curves for the compensated torque at three different $Re_S$ and two radius ratios $\eta=0.357$ (experiments from this study) and $\eta=0.5$ (taken from \cite{Merbold13}) collapse in the interval between $R_{\Omega,opt}(\eta=0.357)$ and $R_\Omega=1$. This corresponds to TC flow between slight counter rotation, and co-rotation up to the Rayleigh stability line. For smaller values of $R_\Omega$, radial partitioning arises and the $\eta$ dependence of $Nu_\omega$ is no longer absent. The statements by \citet{Dub2005} and \citet{Brauckmann2016} hold true even for large values of $R_C$.

\section{Flow topology}
\begin{figure*}
\centering
\includegraphics{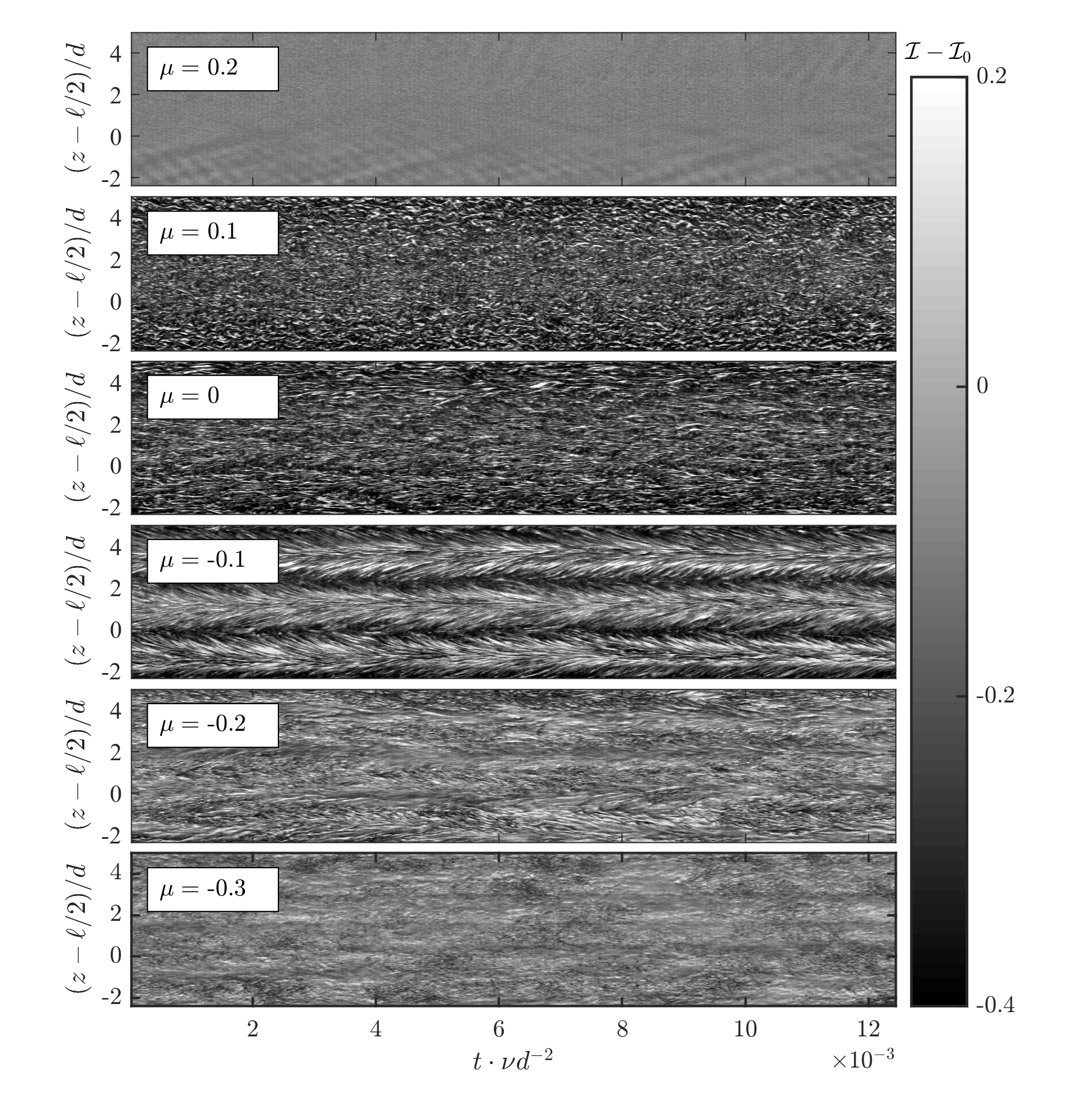}
\caption{Space-time diagrams of the intensity distribution along an axial, central line over time for the shear Reynolds number $Re_S=2.5\times 10^4$ and different $\mu$. The time coordinate is normalized by the viscous time scale $\tau_{vis}=d^2/\nu$ and the axial coordinate by the gap width $d$. All videos have been acquired at 60\,Hz.}
\label{fig:vis_exp}
\end{figure*}

The excellent agreement between our results for the location of the maximum transport of angular momentum and the prediction of \citet{Brauckmann13b} suggests, that the torque maximum is caused by strengthened large-scale Taylor rolls. To verify this explanation, we performed flow visualisations as described in section \ref{sec:exp}. The intensity distribution $\tilde{\mathcal{I}}=\mathcal{I}-\mathcal{I}_0$ of an axial central line for the observed flow as function of time $t$ at the shear Reynolds number $Re_S=2.5\times 10^4$ and different $\mu$ is depicted in Fig.\,\ref{fig:vis_exp}.
\noindent For co-rotation in the Rayleigh stable regime at $\mu=0.2$, the flow is laminar indicated by a homogeneous light intensity over time. However, slightly bright and inclined lines are visible especially below mid height, which are crossing and may be fingerprints of endwall induced Ekman vortices. This would explain the measured Nusselt number of $Nu_\omega \approx1.3$ in that regime. When the rotation ratio is decreased into the linearly unstable regime at $\mu=0.1$, the flow becomes fully turbulent resulting in a strong fluctuation of the light intensity in the axial as well as in the temporal coordinate direction. The flow appears as black background with bright spots randomly distributed in the whole field of view. In case of pure inner cylinder rotation ($\mu=0$), the intensity distribution appears similar to the previous one with small differences for the bright spots. Here, these spots are smeared into the direction of the temporal coordinate direction. When a rotation rate of $\mu=-0.1$ close to $\mu_{\max}$ is adjusted, prominent large-scale vortices are formed inside the flow conforming with the theory of \citet{Brauckmann13b}. Across the axial length of approximately six gap widths, three vortex pairs can be identified , which are stationary in time concerning their axial position. At even higher counter rotation rates, the intensity distributions become again brighter, more homogeneous and the large-scale rolls disappear. This evolution is due to the stabilization of the flow in the outer gap region induced by a stronger rotation of the outer cylinder. As these videos only visualize a small region close to the outer cylinder wall, a probably further existence of the turbulent Taylor vortices in the inner gap can not be shown.

\begin{figure}[htb]
\centering
\includegraphics[trim=0 0 0 260, clip]{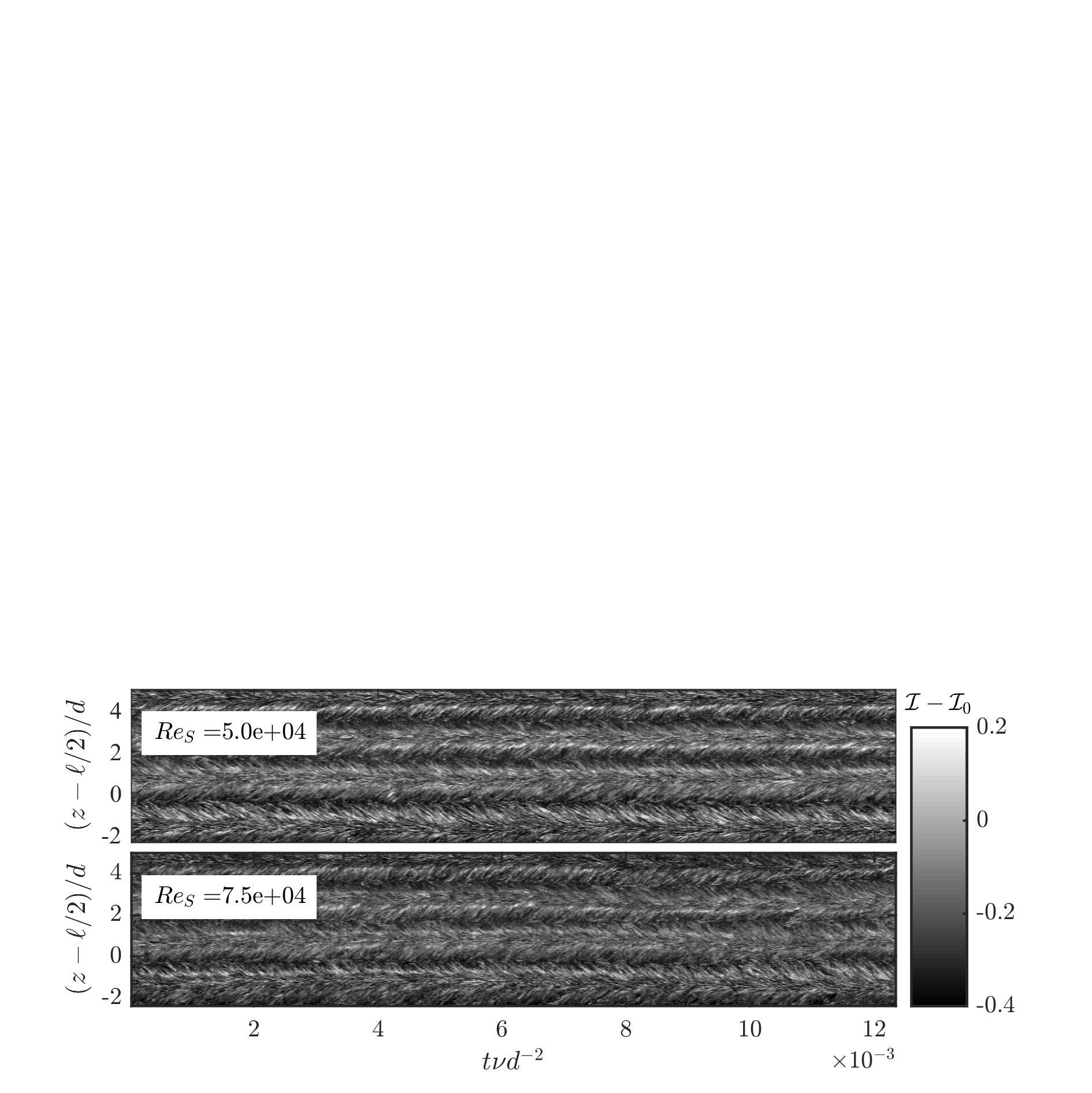}
\caption{Space-time diagrams of the intensity distribution along an axial, central line over time for the shear Reynolds numbers of $Re_S=5\times 10^4$ and $Re_S=7.5\times 10^4$ for $\mu=-0.1$. The time coordinate is normalized by the viscous time scale $\tau_{vis}=d^2/\nu$ and the axial coordinate by the gap width $d$. All videos have been acquired at 60\,Hz.}
\label{fig:vis_exp2}
\end{figure}

We also investigated two additional flow states in the region of the torque maximum at $\mu=-0.1$ for higher shear Reynolds numbers, depicted in Fig.\,\ref{fig:vis_exp2}. Again, we find prominent large-scale rolls in the flow, which are stable in time. In contrast to the flow at $Re_S=2.5\times 10^4$, both cases exhibit four Taylor roll pairs over an axial length of approximately six gap widths. We would like to note, that we have not used a fixed acceleration rate for the cylinders, whose variation can cause different flow states \citep{Martinez2014}. Also, we did not observe a swap from three to four roll pairs or back at a constant shear Reynolds number. Nevertheless, our finding suggests that multiple flow states defined by the overall number of Taylor vortices filling the gap are possible also in this very wide-gap TC geometry. To quantitatively specify the importance of turbulence and large-scale rolls in the flow, we calculate the two following quantities:

\begin{align} \label{Nu_dec}
\begin{split}
\mathcal{I}_{turb} & =\left \langle \left( \mathcal{\tilde{I}}-\left\langle \mathcal{\tilde{I}} \right\rangle_t \right) ^2 \right\rangle_{t,z}, \\
\mathcal{I}_{LSC} & = \sigma_z \left( \left\langle \mathcal{\tilde{I}} \right\rangle_t^2 \right).
\end{split}
\end{align}

\begin{figure}[htb]
\centering
\includegraphics{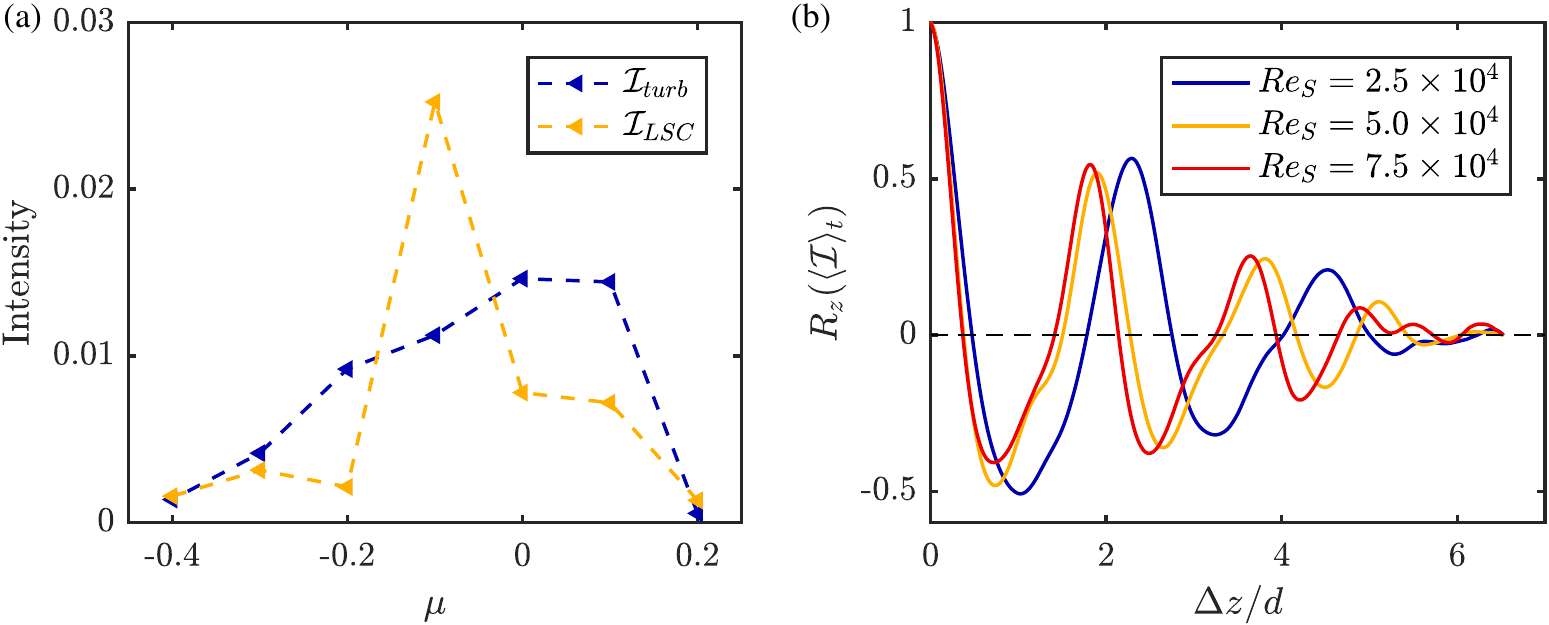}
\caption{(a) Fluctuation of the light intensity $\mathcal{I}_{turb}$ and amplitude of the temporal averaged axial intensity variation $\mathcal{I}_{LSC}$ for $Re_S=2.5\times 10^4$ as function of $\mu$. (b) Axial autocorrelation coefficient of the temporal averaged axial intensity profile for $\mu=-0.1$.}
\label{fig:fractions}
\end{figure}

\noindent $\mathcal{I}_{turb}$ is a measure for the fluctuations of the light intensity and therefore for the turbulence inside the flow. Beside, $\mathcal{I}_{LSC}$ quantifies the amplitude of the temporal averaged axial intensity variation and therefore increases, when large-scale rolls are formed in the gap. Both quantities are depicted in Fig.\,\ref{fig:fractions}a as function of the rotation ratio $\mu$. 

The intensity fluctuation $\mathcal{I}_{turb}$ strongly increase, when the rotation ratio is reduced starting at $\mu=0.2$ in the Rayleigh stable regime into the unstable regime at $\mu=0.1$. For a further decrease of $\mu$, $\mathcal{I}_{turb}$ continuously decreases up to $\mu=-0.4$. On the other hand, $\mathcal{I}_{LSC}$ increases from $\mu=0.2$ to $\mu=-0.1$. For smaller $\mu$, a rapid breakdown of the amplitude is seen. Accordingly, the results of visual inspection of the space-time diagrams depicted in Fig.\,\ref{fig:vis_exp} are supported quantitatively by the parameters $\mathcal{I}_{turb}$ and $\mathcal{I}_{LSC}$. Moreover, in Fig.\,\ref{fig:fractions}b, the axial autocorrelation function $R_{z}$ of the temporally averaged axial intensity profile is depicted for the three flow cases at $\mu=-0.1$ in the region of the torque maximum. The correlation coefficients depict a prominent oscillation in the axial coordinate direction due to the large-scale Taylor rolls and the first minimum is a measure for the axial wavelength $\lambda$. We find $\lambda(Re_S=2.5\times 10^4)=1.02d$, $\lambda(Re_S=5.0\times 10^4)=0.73d$ and $\lambda(Re_S=7.5\times 10^4)=0.74d$. For the lowest shear Reynolds number, the Taylor vortices capture nearly one gap width in the axial direction, but become axially compressed at higher $Re_S$. The effect of changes in the vortex wavelength on the angular momentum transport has been investigated in more detail for example by \citet{Martinez2014} and \citet{Ostilla14b}.

\newpage

\begin{figure}[htb]
\centering
\includegraphics[trim=0cm 0 0cm 200, clip]{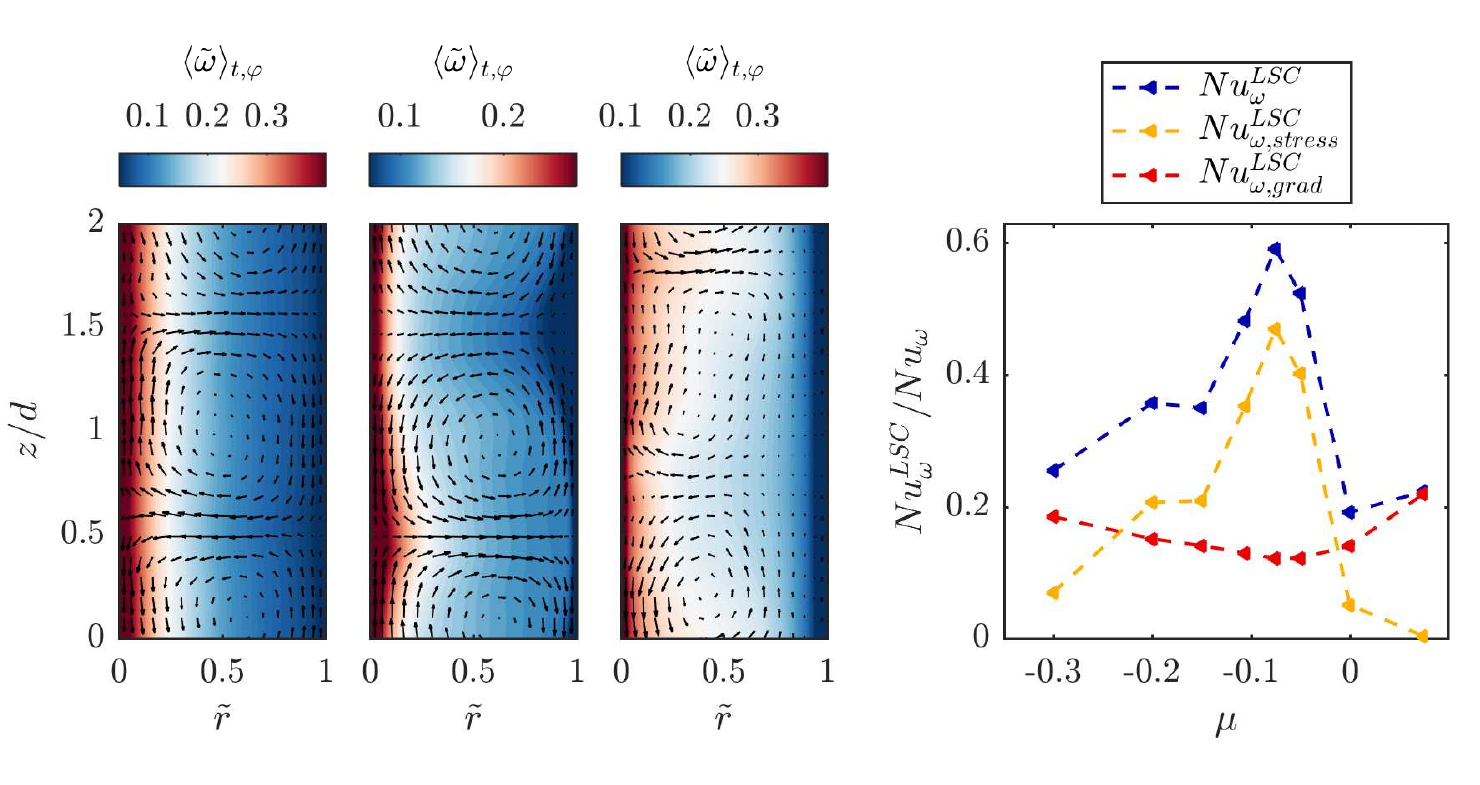}
\caption{Numerically determined azimuthally and temporally averaged angular velocity for $Re_S=2\times10^4$ and $\mu=0$ (left), $\mu=-0.08$ (middle) and $\mu=-0.3$ (right). Arrows representing the average radial and axial velocities are superimposed. Right next to it, the contribution of the large-scale circulation to the overall Nusselt number as function of $\mu$ is shown for the same driving.}
\label{fig:Lme_Mus}
\end{figure}

In simulations, as we have fixed the axial periodicity to a relatively low number ($\Gamma=2$), so we cannot expect to see switching between roll wavelengths. Due to the very demanding requirements in the axial direction, confirmation of these findings in numerics requires a large amount of computational resources. However, we can visualize the footprint of the rolls in the mean angular velocity fields and narrow down more precisely the $\mu$-range, where Taylor vortices exist. The results are shown in Fig. \ref{fig:Lme_Mus} for $Re_S=2\times10^4$ and different rotation ratios $\mu$. For pure inner cylinder rotation, while there is a structure which can be seen in the average radial and axial velocities, no footprint is observed in the average angular velocity. For $\mu=-0.08$, we see the roll, but now also see a footprint on the average angular velocity, confirming the experimental findings. Finally, for $\mu=-0.3$, the stabilization coming from the outer cylinder is enough to distort the rolls, similar to what was seen in Ref. \cite{Ostilla14a} for $\eta=0.714$. To determine the $\mu$-range, where Taylor vortices exist, we calculate the fraction of the large-scale circulation $Nu_\omega^{LSC}$ on the total momentum transport as it was done in Ref. \citep{Brauckmann13b} based on the flow field decomposition $u=\bar{u}+u^\prime$ with $\bar{u}=\langle u \rangle_{\varphi,t}$:

\begin{equation}
Nu_\omega^{LSC}=\underbrace{\langle r^3 \langle \bar{u}_r \bar{\omega} \rangle_{\varphi,z,t} \rangle_r/J_\omega^{lam}}_{Nu_{\omega,stress}^{LSC}}- \underbrace{\langle r^3 \nu \partial_r \langle \bar{\omega} \rangle_{\varphi,z,t} \rangle_r/J_\omega^{lam}}_{Nu_{\omega,grad}^{LSC}}
\end{equation}

\noindent The gradient fraction of ${Nu}_\omega^{LSC}$ is larger than zero independent of the existence of the large-scale rolls (see Fig. \ref{fig:Lme_Mus}), which is why we focus here on the stress fraction $Nu_{\omega,stress}^{LSC}$. This quantity clarifies that Taylor vortices contribute considerably to the angular momentum transport in the range of $-0.3 < \mu < 0$, and in the region of the torque maximum their contribution is around 60\%.

\section{conclusion}
We have investigated experimentally and numerically the angular momentum transport and the corresponding flow structure in a fully turbulent Taylor-Couette flow at a radius ratio of $\eta=0.357$. For an outer cylinder at rest, no pure power law scaling is found across the $Re_s$ range explored and the effective scaling exponent $\alpha$ varies noticeable in the region of $1.3\times 10^4 \leq Re_S\leq 4\times 10^4$. Interestingly, the frictional Reynolds number at the inner cylinder at the lower end of this region is nearly identical to the outer frictional Reynolds number at the upper end: both are approximately $Re_{\tau,crit}\approx 230$.
At first glance, this points towards a boundary layer transition that triggers the ultimate regime. However, this is ruled out by two facts: (\emph{i}) The calculated effective scaling exponent throughout this region is smaller than the predicted lower limit of $\alpha_{marg}=5/3$ according to \citet{King1984} and \citet{Marcus1984}, and (\emph{ii}) non-logarithmic boundary layer profiles at the outer wall. Instead, our investigations should be assigned to the classical-turbulent regime. By exploring the bulk properties, we found that the transition itself is associated with the capacity of the outer cylinder to emit small-scale plumes. It ends as the angular momentum profile in the bulk reaches the condition of marginal stability, i.e a flat profile at approximately the arithmetic mean $\tilde{L}=0.5$ of both cylinders. 
 
\begin{figure}[htb]
\centering
\includegraphics{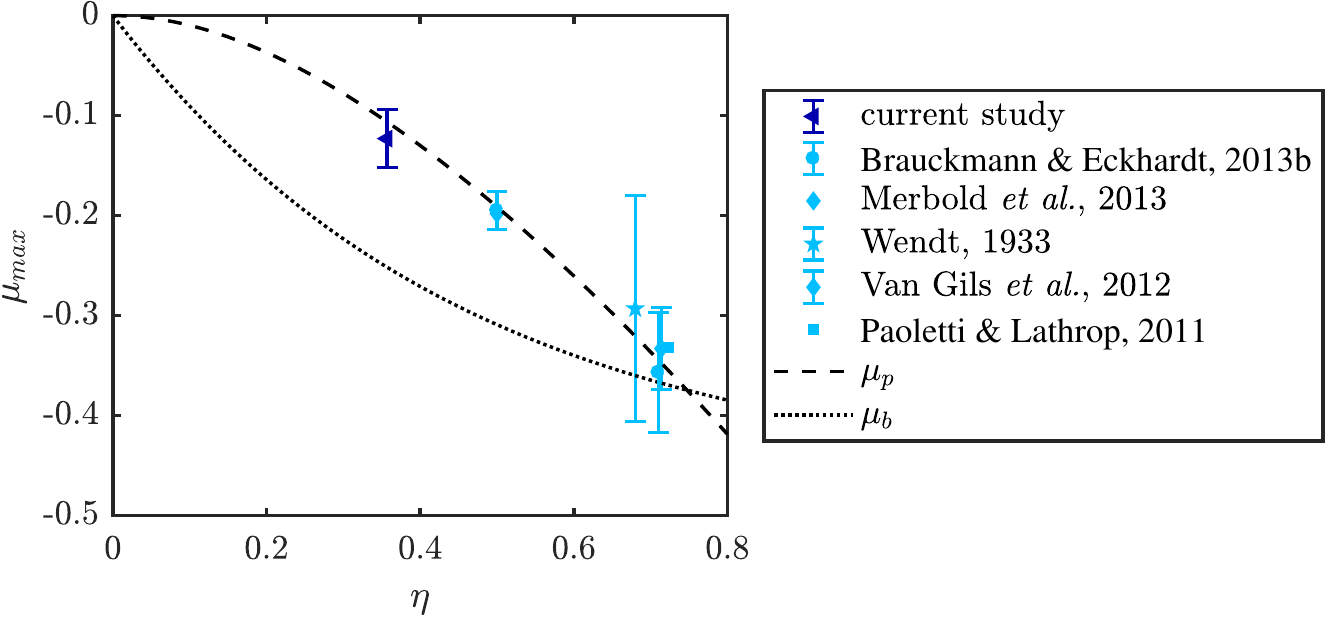}
\caption{Location of the torque maximum rotation rate $\mu_{\max}$ of our study in comparison to different results for smaller gap TC flows, restricted to medium gaps with $\eta \leq 0.8$. The dashed and dotted lines represent the predictions of the torque maximum according to \citet{Brauckmann13b} and \citet{VanGils2012}, respectively.}
\label{fig:vgl_mumax}
\end{figure}

In case of independently rotating cylinders, a maximum in torque was found at $\mu_{\max}(\eta=0.357)=-0.123 \pm 0.030$, which is induced by the formation and strenghtening of large-scale vortices. This value of $\mu_{\max}$ and the mechanism are in line with Ref. \citep{Brauckmann13b}. The state-of-the-art results of the torque maximum locations for medium and wide gaps are summarized in Fig. \ref{fig:vgl_mumax} together with our results. The good agreement with the prediction of \citet{Brauckmann13b} demonstrates, that the physical mechanism for $\mu_{\max}$ does not change in that $\eta$-regime. Indeed, we found that as long as radial partitioning does not appear, the Nusselt number for $\eta=0.357$ was in good agreement to those found for $\eta\geq 0.5$, as proposed by \citep{Brauckmann2016b}. Finally, two different wavelengths of large-scale vortices have been discovered for $Re_S>Re_{S,crit,1}$ at different forcings. In order to explore the characteristics of such multiple flow states in wide-gap TC flows in more detail, further investigations in the spirit of those of Refs. \citep{Huisman2014,vanderVeen2016} carried out at smaller gaps are needed.  

\section*{Acknowledgements}
We gratefully acknowledge financial support by the Deutsche Forschungsgesellschaft (DFG) under Grant numbers EG100/15-2 and EG100/23-1.

\appendix

\begin{table}[tp]
\caption{\label{tab:dnsresults} Full details of the numerical resolutions used and the pseudo-Nusselt number $Nu_\omega$ and inner cylinder frictional Reynolds number $Re_{\tau,i}$ obtained from each of them. }
\begin{tabular}{cccccccc}
  \hline
  $Re_s$ & $\mu$ & $R_\Omega$ & $N_\theta$ & $N_r$ & $N_z$ & $Nu_\omega$  & $Re_{\tau,i}$ \\
 \hline
$5\times10^3$ & 0      & 0.643 & 256 &  256 &  512 &  4.55 & 120  \\
$7.07\times10^3$ & 0      & 0.643 & 256 &  256 &  512 &  5.49 & 158 \\ 
$1\times10^4$ & 0.075  & 0.841 & 256 &  256 &  512 &  3.76 & 154    \\
$1\times10^4$ & 0      & 0.643 & 256 &  256 &  512 &  6.7  & 209    \\
$1\times10^4$ & -0.05  & 0.527 & 256 &  256 &  512 &  8.45 & 236    \\
$1\times10^4$ & -0.075 & 0.472 & 256 &  256 &  512 &  8.67 & 238    \\
$1\times10^4$ & -0.1   & 0.408 & 256 &  256 &  512 &  8.45 & 235    \\
$1\times10^4$ & -0.15  & 0.324 & 256 &  256 &  512 &  8.15 & 232    \\
$1\times10^4$ & -0.2   & 0.236 & 256 &  256 &  512 &  7.91 & 229    \\
$1\times10^4$ & -0.3   & 0.079 & 256 &  256 &  512 &  6.99  &214   \\
$1.41\times10^4$ & 0      & 0.643 & 256 &  256 &  512 &  8.16 &276   \\ 
$2\times10^4$ & 0.075  & 0.841 & 384 &  256 &  512 &  5.54  &264   \\
$2\times10^4$ & 0      & 0.643 & 384 &  256 &  512 &  10.1  & 363   \\
$2\times10^4$ & -0.05  & 0.527 & 384 &  256 &  512 &  12.6  & 410   \\
$2\times10^4$ & -0.075 & 0.472 & 384 &  256 &  512 &  13.5  & 421   \\
$2\times10^4$ & -0.1   & 0.408 & 384 &  256 &  512 &  13.4  & 417   \\
$2\times10^4$ & -0.15  & 0.324 & 384 &  256 &  512 &  13.1  & 413   \\
$2\times10^4$ & -0.2   & 0.236 & 384 &  256 &  512 &  12.5  &404  \\ 
$2\times10^4$ & -0.3   & 0.079 & 384 &  256 &  512 &  11.1  & 378\\   
$4\times10^4$ & 0      & 0.643 & 768 &  384 &  768 &  15.4  & 635 \\
  \hline
\end{tabular}
\end{table}

\clearpage

\bibliography{main}
\end{document}